\documentclass[final,5p,twocolumn]{elsarticle}

\pdfoutput=1 
\input{_preambles/preamble_elsarticle.tex}
\usepackage[usenames,dvipsnames]{xcolor}

\usepackage[utf8]{inputenc}
\usepackage{mathtools}
\usepackage{amsmath}    
\usepackage{pifont}     
\usepackage{booktabs} 

\usepackage{fancyvrb}   

\newcommand{\OMIT}[1]{} 

\newcommand{\R}{\textsf{R}}               
\newcommand{\CPP}{\textsf{C}{\tt ++}} 

\newcommand{\mtext}[1]{\noindent\colorbox{yellow}{\parbox{0.98\linewidth}{#1}}}%

\definecolor{Gray}{rgb}{0.9,0.9,0.9}

\definecolor{shadecolor}{RGB}{220,220,220}

\begin{document}

\begin{frontmatter}

\title{
Asymptotic Experiments with Data Structures:\\ 
Bipartite Graph Matchings and Covers
}


\author{Eason~Li}
\ead{yli223@ncsu.edu}
\author{Franc~Brglez}
\ead{brglez@ncsu.edu}
\address{Computer Science, NC State University, Raleigh, NC 27695, USA}
%
%
\begin{abstract}
We consider {\em instances of bipartite  graphs}  and a number of asymptotic performance experiments in three projects: 
(1) top movie lists, 
(2) maximum matchings, and 
(3) minimum set covers.
Experiments are designed to measure the asymptotic runtime performance of abstract data types (ADTs) in three programming languages: Java, \R{}, and C++.
The outcomes of these experiments may be surprising.
In project (1), the best ADT in \R{} consistently outperforms all ADTs in public domain Java libraries, including the library from Google. 
The largest movie list has $2^{20}$ titles.
In project (2), the Ford-Fulkerson algorithm implementation in \R{} significantly outperforms Java.
The hardest instance has 88452 rows and 729 columns.
In project (3),  
a stochastic version of a greedy algorithm 
in R can significantly outperform a state-of-the-art stochastic
solver in C++ on instances with 
$num\_rows \ge 300$ and $num\_columns \ge 3000$. 
\end{abstract}
\begin{keyword}
\texttt{ADTs in Java, R, and C++}        \sep 
\texttt{bipartite graphs}                \sep 
\texttt{maximum matchings}               \sep 
\texttt{minimum set covers}              \sep 
\texttt{runtime performance experiments} \sep
\texttt{asymptotic complexity}   
\end{keyword}

\end{frontmatter}
 
\section{Introduction}  \label{sec_introduction}
\noindent
The title of this article was by inspired by a 2-sentence abstract from a   92-page 
publication~\cite{OPUS-matching-1998-Fueredi-bigraphs}: 
{\it ``Almost all combinatorial questions can be reformulated as either a matching or a covering problem of a hypergraph. In this paper we survey some of the important results.''.}

Rather than  theorems and proofs, 
this article is about {\it asymptotic performance  experiments}
on matching and covering problems with data structures that represent the hypergraph as a bipartite graph: a matrix with $m$ rows and $n$ columns. For an illustration of matching and covering problems
addressed in this article, see the example of the 11-row, 9-column bigraph in Section~\ref{sec_matching}, Figure~\ref{fg_bgmc_matching_cover}.

Companion articles~\cite{OPUS2-2022-coupon-arxiv-Brglez, OPUS2-2022-mclass-arxiv-Brglez} 
provide the background and the motivation
for a series of experiments we report in four sections of this article:
\begin{description}

\item[\sf{Beyond CSC316 and Java}]~\\\
Data structures impact the 
asymptotic runtime performance when creating lists
such as {\it the top 10 most frequently watched movies}.
The key finding is that the {\tt data.table} structure in \R~\cite{OPUS-R-manual} 
significantly outperforms all of the best-known and widely
available ADTs in Java. The largest movie list has $2^{20}$ titles.

\item[\sf{Maximum Matching in Bipartite Graphs}]~\\\
We compare runtime performance of two 
public-domain implementations of the Ford-Fulkerson algorithm: Java and \R{}.
The hardest instance has 88452 rows and 729 columns.
Again, \R{} significantly outperforms the implementation in Java.

\item[\sf{Greedy Heuristic Distributions for Set Cover}]~\\\
The importance of greedy heuristics is increasing as the
instance sizes increase for problems such as {\em the minimum set cover}.
We demonstrate that a stochastic version of a greedy algorithm 
in \R{} can significantly outperform a state-of-the-art stochastic
solver in C++ on instances with 
$num\_rows \ge 300$ and $num\_columns \ge 3000$. 

\item[\sf{Future Work}]~\\\
The work in progress includes extensions of new heuristics,
outlined in the  companion article~\cite{OPUS2-2022-mclass-arxiv-Brglez},
to a number of {\it hard} combinatorial optimization problems.

\end{description}

\begin{figure*}[t!]
\vspace*{-4ex}
\centering
%
%
%
%
%
%
%

\includegraphics[width=1.00\textwidth]{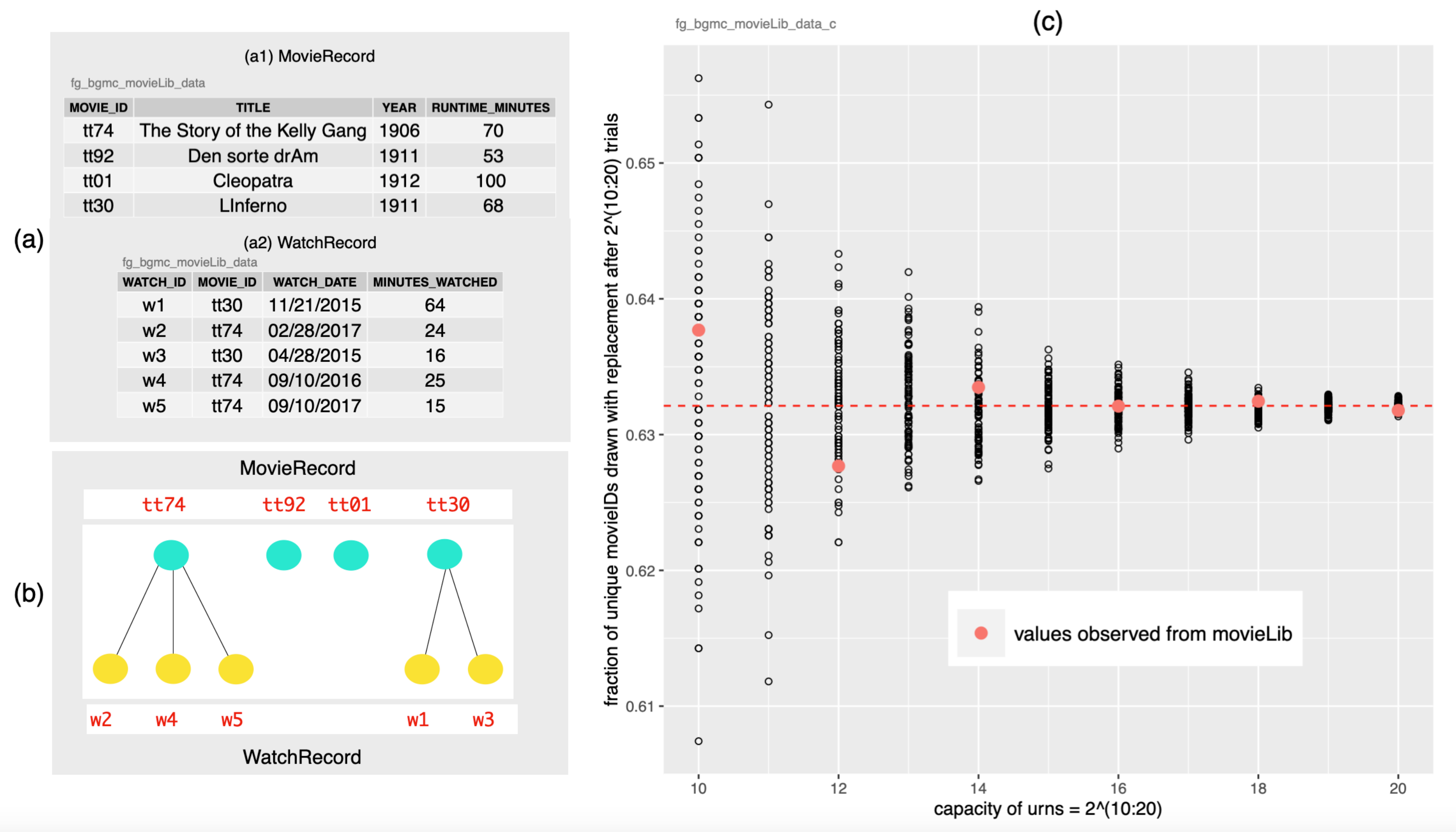}

\caption{Representations of data sets used in the performance experiments with 
the {\it MovieLib}   
data set introduced in Section~\ref{sec_316}:
{\sf (a)}~a tabular organization of files {\it MovieRecord} and  {\it WatchRecord}, 
{\sf (b)}~a bigraph that illustrates relationships between the
items from the two files in (a), and
{\sf (c)}~a statistical model, based on $2^{10}~...~2^{20}$ trials of sampling with replacement   
from 11 urns, each with urn capacity of holding $2^{10}~...~2^{20}$ unique  {\it movieID} tags.
The actual {\it MovieLib}  is represented with 6 urns: their sizes are
$2^{10}$,  $2^{12}$,  $2^{14}$, $2^{16}$, $2^{18}$,  and $2^{10}$.
Notably, the fraction of unique  {\it movieID} tags observed by analyzing actual data from  {\it MovieLib} 
is also converging towards $1 - e^{-1} = 0.6321$ and
is well within the expected range for this experiment.
}

\label{fg_bgmc_movieLib_data}
\end{figure*}
\vspace*{-4ex}

\section{Beyond CSC316 and Java} \label{sec_316}
\noindent
%
CSC316 is a junior-level course in data
structures and algorithms~\cite{OPUS-csc316-fall-2020}.
A class project relevant to this article, {\tt PackFlix}, 
explored the impact of data structures on runtime performance. 
The data for  {\tt PackFlix}, modified for educational purposes, originated with IMDb~\cite{OPUS-csc316-dataset}. 
The project objective was to not only analyze the watching histories
of customers by designing a software prototype {\tt PackFlix}
in Java. The primary objective was to study the 
impact of data structures on the 
asymptotic runtime performance to create lists
such as {\it the top 10 most frequently watched movies}.
The primary input to {\tt PackFlix} is a directory path to 
 {\tt movieLib} which contains file pairs of increasing size:
{\tt movieRecords} and {\tt watchRecords}.

The example in Figure~\ref{fg_bgmc_movieLib_data} illustrates the organization 
of two data files and the range of
file sizes that are being considered for the experiments. 
 
Two tables in Figure~\ref{fg_bgmc_movieLib_data}a,
{\it MovieRecord} and  {\it WatchRecord} are related.
Columns in the first table refer to a unique movieID, a title, a release year, and runtime in minutes.
Columns in the second  table refer to a watchID, movieID, a watch date, and minutes watched.

The Figure~\ref{fg_bgmc_movieLib_data}b is a bipartite graph (a bigraph) 
that illustrates relationships between the items from the two files 
in Figure~\ref{fg_bgmc_movieLib_data}a.
The movie {\it tt74} has been watched 3 times, the movie {\it tt30}  has been watched once, 
and the remaining two movies have not been watched. 
Clearly, the most popular movie is {\it tt74}.

The Figure~\ref{fg_bgmc_movieLib_data}c depicts experiments with a series of 
{\em urn models}~\cite{OPUS-book_stats-1977-Wiley-Johnson-urn_models}, 
based on $2^{10}~...~2^{20}$ trials of sampling with replacement   
from 11 urns, each holding $2^{10}~...~2^{20}$ unique  {\it movieID} tags.
The experiments are structured to measure the ratio of unique {\it movieID} tags observed after
$2^{10}~...~2^{20}$ trials. 
As the size of urns and the number of trials increases, this ratio converges to
the value of $1 - e^{-1} = 0.6321$.
The movies that are actually catalogued in {\it MovieLib}  are represented as six urns: their sizes are
$2^{10}$,  $2^{12}$,  $2^{14}$, $2^{16}$, $2^{18}$,  and $2^{20}$.
 Notably, both the models and the analysis of
actual data from  {\it MovieLib} converge to the expected value of $1 - e^{-1} = 0.6321$.

\begin{figure*}[t!]
\centering
%
%
%
%
%
%
%
%

\includegraphics[width=1.0\textwidth]{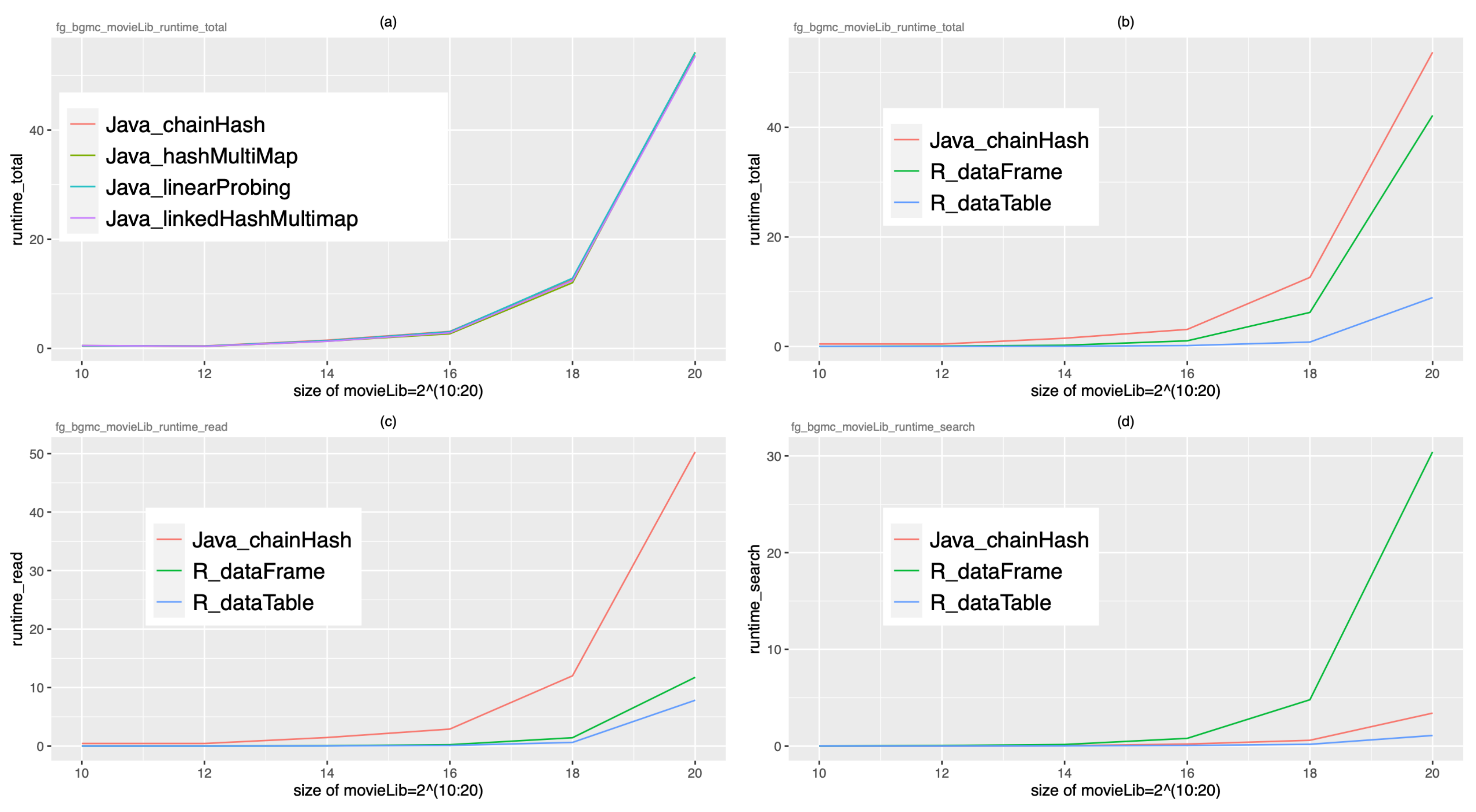}

\caption{
Asymptotic runtime performance experiments with instances from {\it movieLib}, 
based on Java and R code. In each case, the objective is to retrieve the top 10
movies after reading two sets of files: one listing movies and one listing 
viewer interactions with each movie.
In (a), we report {\it runtime\_total} for best four data structures in Java; 
a model used in CSC316 class.
In (b), we report on {\it runtime\_total} for the single best data structure in Java ({\t chainHash}) in comparison with two best data structures in R ({\t dataFrame} and {\t dataTable}).
In (c), we report {\it runtime\_read} for {\it chainHash} in Java and {\it dataFrame} and {\it dataTable}
in R.
In (d), we report {\it runtime\_search} for {\it chainHash} in Java and {\it dataFrame} and {\it dataTable}
in R.
There is no doubt that, for  instances from {\it movieLib}, 
R significantly outperforms Java -- with {\it dataTable} an asymptotically better data structure in 
comparison with {\it dataFrame}.
}
\OMIT{
\caption{Experiments that analyze the total runtime, reading runtime and searching runtime for all Java and R programs. Data table version in R has a significantly improvement in runtime performance comparing to the others. The main obstacle that slows down the Java program is when reading data files and converting each line of information into an object. However, the Java program runs relatively fast when manipulating in the movieLib. Overall, the data table version of R program outperforms in any aspect including reading files and searching movies.}
}
\label{fg_bgmc_movieLib_runtime}
\end{figure*}

\subsection{{\sf Data Structures and Java Libraries}}
\noindent
Data structures introduced in CSC316 are standard Java libraries introducing a number of Java ADTs, from {\it Linked List} to {\it Linear Probing Hash Map}.

\par\vspace*{0.9ex}
In this article, we extend our runtime performance experiments to additional Java ADTs:
{\it Hash MultiMap} and {\it Linked Hash MultiMap} from
{\it Google Guava}~\cite{OPUS-csc316-guava} and  {\it Chain Hash Map} from
{\it net.datastructures}, posted at the Brown University~\cite{OPUS-csc316-net-datastructure}.

\par\vspace*{0.9ex}
The Java code uses Map ADT to pair each key and value.
Initially, our R code also paired each key with a value using a hash function. 
However, the runtime performance was worse than Java ADT.
This led to exploration of two data structures in R:
{\tt data.frame}~\cite{OPUS-csc316-data-frame} and 
{\tt data.table}~\cite{OPUS-csc316-data-table}. 
The {\it Eureka moment} came with the observation that  
{\tt data.table} in R significantly
outperforms the best Java version. For details,
see Figure~\ref{fg_bgmc_movieLib_runtime}.
\par\vspace*{0.9ex}

\subsection{{\it \sf Runtime experiments: Java vs R}}
\noindent
The four plots in Figure~\ref{fg_bgmc_movieLib_runtime},
illustrate the runtime performance for {\tt PackFlix} in Java and R. 
In the previous section, we introduced four Map ADTs in Java, 
which are from course work and public domain. 

\begin{description}

\item[\sf{Plot in Figure~\ref{fg_bgmc_movieLib_runtime}a}]~\\\
is a repeat of  experiments in CSC316:
it depicts {\tt runtime\_total} of {\tt PackFlix} with the Map ADTs from Java. 
Results show that these runtimes are statistically equivalent. 
For the follow-up experiments in
Figures~\ref{fg_bgmc_movieLib_runtime}b,c,d
we select 
{\tt Java\_chainHash} as a representative of the best Java ADTs to be compared with the two  ADTs in R.

\item[\sf{Plot in Figure~\ref{fg_bgmc_movieLib_runtime}b}]~\\\
depicts {\tt runtime\_total} of {\tt PackFlix} with 
{\tt Java\_chainHash},
{\tt R\_dataFrame}, and 
{\tt R\_dataTable}. 
Here, we observe that {\tt runtime\_read} of {\tt PackFlix} under Java is 
significantly outperformed by {\it both} ADTs in R.
Questions that arise are these:\\
(1) Which ADT is the best when reading files and initializing the respective data structures?\\
(2) Which ADT is the best when searching the dataset before returning the top 10 movies?

\item[\sf{Plot in Figure~\ref{fg_bgmc_movieLib_runtime}c}]~\\\
depicts only the {\tt runtime\_read} of {\tt PackFlix} with 
{\tt Java\_chainHash},
{\tt R\_dataFrame}, and 
{\tt R\_dataTable}. 
Again, we observe that {\tt runtime\_read} of {\tt PackFlix} under Java is 
significantly outperformed by {\it both} ADTs in R.
In principle, Java can read large datasets efficiently. 
However, in {\tt PackFlix}, it not only needs to read line by line from each data files, 
but it also needs to convert each line of data into objects and save them into a global array. This appears as the major factor that Java programs in {\tt PackFlix} cannot compete with ADTs in R. 
This question best left to R developers:
why does {\tt R\_dataTable}, under  {\tt runtime\_read},
start to outperform {\tt R\_dataFrame} at 
instance sizes $\ge 2^{18}$?

\item[\sf{Plot in Figure~\ref{fg_bgmc_movieLib_runtime}d}]~\\\
focuses on the {\tt runtime\_search} of {\tt PackFlix} with 
{\tt Java\_chainHash},
{\tt R\_dataFrame}, and 
{\tt R\_dataTable}. 
All of these programs return the same list of 
{\it the top 10 most frequently watched movies}.
However, there are significant differences 
in the {\tt runtime\_search} 
for the largest movie list with $2^{20}$ titles.
Again, {\tt R\_dataTable} significantly outperforms
{\tt Java\_chainHash}. But here, {\tt Java\_chainHash}
significantly outperforms {\tt R\_dataFrame}.
Another question for developers of R:
why does the gap in {\tt runtime\_search} between {\tt R\_dataFrame}
and {\tt R\_dataTable} increases so rapidly 
for {\em this dataset}?


\end{description}

\subsection{Correlation Experiments with {\tt MovieLib} Dataset}
\noindent
We complete the analysis of experiments with {\tt PackFlix} and 
the {\tt MovieLib} dataset by a frequency analysis of the 
top 10 movie watched and the total number of movies watched 
for the largest movie list with $2^{20}$ titles.

\begin{figure}[h!]
\vspace*{-2ex}
\centering

\includegraphics[width=0.48\textwidth]{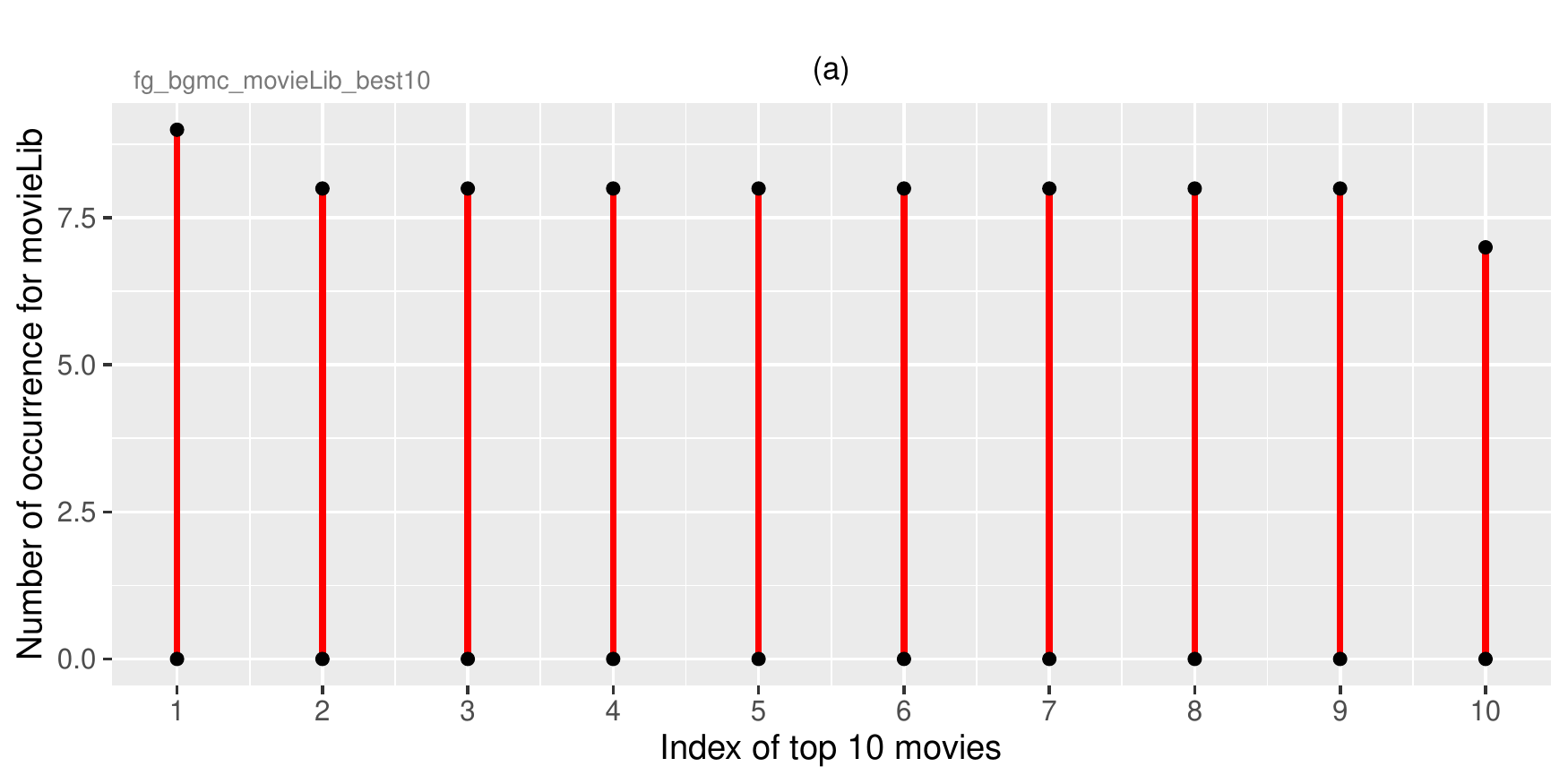}
\\[2ex]
\includegraphics[width=0.48\textwidth]{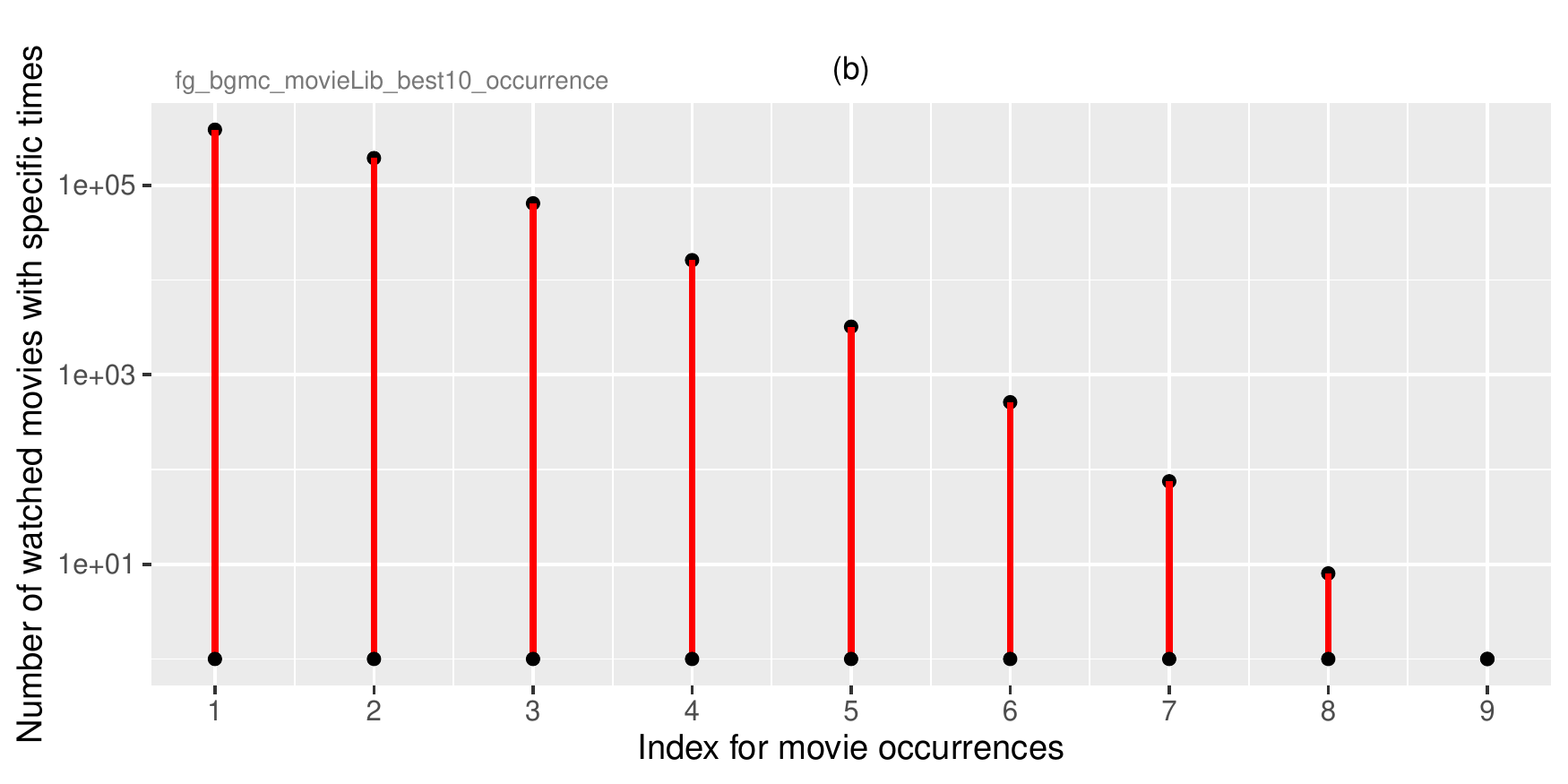}

\caption{
These two plots relate results for the largest movie list with $2^{20}$ titles.
The plot (a) depicts the frequency of top 10 movies watched. For example, the movie with index 1 has been watched 9 times, movies with indices 2-9 have been watched 8 times, etc. 
\\
The plot (b) counts the total number of movies watched. For example, 385128 movies have been watched only once (index = 1), 75 movies have been watched 7 times (index = 7), only one movie has been watched 9 times (index = 9).
\vspace*{-3ex}
}
\label{fg_bgmc_movieLib_best10}
\end{figure}

\begin{description}

\item[\sf{Plot in Figure~\ref{fg_bgmc_movieLib_best10}a}]~\\\
depicts the frequency of top 10 movies watched. 
The movie with index 1 has been watched 9 times, movies with indices 2-9 have been watched 8 times, etc.

\item[\sf{Plot in Figure~\ref{fg_bgmc_movieLib_best10}b}]~\\\
counts the total number of movies watched: 
385,128 movies have been watched only once (index = 1), 75 movies have been watched 7 times (index = 7), only one movie has been watched 9 times (index = 9).

\end{description}
%
%
%
%
%

\section{Maximum Matching in Bipartite Graphs} \label{sec_matching}
\noindent
%
%
Given a graph $G=(V,E)$, a matching M in G is a set of pairwise non-adjacent edges.
A maximum matching, also known as maximum-cardinality matching,
is a matching that contains the largest possible number of edges. 
Every maximum matching is maximal, but not every maximal matching 
is a maximum matching.

Versions of maximum matching problems arise in a number of contexts
and applications:
from flow and neural networks, scheduling and planning, modeling bonds 
in chemistry, graph coloring, the stable marriage problem, 
to matching kidney donors to kidney donor recipients, etc.

The 59-page chapter on maximum-flow problem formulations 
in~\cite{OPUS-book_algs-1990-MCGraw_Hill-Cormen} includes a section on the
maximum bipartite matching.
Maximum matching runtime in an undirected bipartite graph $G=(V, E)$
ranges from polynomial in $|V|$ and $|E|$ with the  
Ford-Fulkerson method~\cite{OPUS-matching-1956-Math-Ford_Fulkerson-max_flows}
%
to  $O(\sqrt(|V|) |E|)$ with the 
Hopcroft and Karp algorithm~\cite{OPUS-matching-1973-SIAM-Hopcroft_Karp}. 

Computational experiments with maximum bipartite matching in this article are conducted with two solvers that both rely on Ford-Fulkerson method:
one implemented in Java~\cite{OPUS-matching-2021-geeksforgeeks-java}, 
the other implemented in \R{}~\cite{OPUS-matching-2021-igraph-R}.

The bigraph instances in these experiments are the same ones we use for 
the experiments in the next section where we search for the minimum set cover.
The instances have been assembled as larger  instance subsets from variety of 
sources:
the subset of steiner3 instances~\cite{OPUS-setc-2021-Resende-steiner3_data},
the subset of OR-library instances~\cite{OPUS-setc-2014-orlib-Beasley},
and the subset of logic optimization instances~\cite{OPUS2-1993-benchm-Logic_synthesis}.
We converted all files to
the {\it DIMACS cnf format}~\cite{OPUS-cnf-2021-wiki}
with minor extensions.
This format unifies the formulations
of both the  {\it minimum unate} as well as the {\it minimum binate} covering
problems~\cite{OPUS2-2005-cover-DAC-Li}.
The file extension {\tt .cnfU} implies a unate set instance
with {\it unit weights}, the file extension {\tt .cnfW} implies a unate 
or a binate set instance with {\it non-unit weights}.

\begin{table*}[t!]  
  \vspace*{-6ex}
  \centering
\caption{
This table supports discussions in Section~\ref{sec_matching} 
{\it as well as} in Section~\ref{sec_cover}.
Columns denote the name of the instance file, 
number of instance columns ({\it nCols}), 
number of instance rows ({\it mRows}), 
matrix density column  
({\it mDens $:=$ numEdges$/$(nCols*mRows)}), 
maximum matrix degree column  ({\it mCD}), 
a column with relative values of maximum matchings for each instance, 
({\it mP $:=$ max\_matchingSize$/$nCols}), 
best-known-values of the minimum set cover ({\it BKV}), 
Chvatal's upper bound on the minimum set cover ({\it UB}), 
observed set cover statistics with the Chvatal's greedy algorithm ({\it value\_Chvatal\_stats}), and
set cover statistics normalized with {\it BKV}, ({\it BKV\_ratio\_stats}).
Instance file name extensions, {\it .cnfU} and {\it .cnfW}, denote instances with unit weighted columns
and pre-assigned weighted columns, respectively. 
The reported statistics represents values of {\it minimum, median, mean, standard deviation, and maximum}.
Except for instances that are prefixed with {\it **},
the reported statistics are based on experiments with 1000 replications.
Experiments with six instances prefixed with {\it **} are based on
10,000 replications.
}\label{tb_bgmc_data}

  \small{
  \hspace*{-0em}
  \begin{tabular}{l@{\hspace{1\tabcolsep}}c@{\hspace{1\tabcolsep}}c@{\hspace{1\tabcolsep}}c@{\hspace{1\tabcolsep}}c@{\hspace{1\tabcolsep}}c@{\hspace{1\tabcolsep}}c@{\hspace{1\tabcolsep}}c@{\hspace{1\tabcolsep}}l@{\hspace{1\tabcolsep}}c} 
    \toprule
    instance & nCols & mRows & mDens & mCD & mP & BKV & UB & value\_Chvatal\_stats  & BKV\_ratio\_stats\\
    \midrule
    {\bf steiner3} &  &  &  &  & &  & & &\\ 
 s3\_027\_117.cnfU & 27 & 117 & 0.1111 & 13 & 1.00 & 18 & 57.24 & 19,19,19,0,19 & 1.06,1.06,1.06,0.00,1.06 \\ 
 s3\_045\_330.cnfU & 45 & 330 & 0.0667 & 22 & 1.00 & 30 & 110.72 & 31,32,31.87,0.9,33 & 1.03,1.07,1.06,0.03,1.10 \\ 
 s3\_081\_1080.cnfU & 81 & 1080 & 0.0370 &  40 & 1.00 & 61 & 260.99 & 65,65,65,0,65 & 1.07,1.07,1.07,0.00,1.07 \\ 
 s3\_135\_3015.cnfU & 135 & 3015 & 0.0222 &  67 & 1.00 & 103 & 493.3 & 107,107,107.92,1.35,111 & 1.04,1.04,1.05,0.01,1.08 \\ 
 s3\_243\_9801.cnfU & 243 & 9801 & 0.0123 &  121 & 1.00 & 198 & 1064.67 & 211,211,211,0,211 & 1.07,1.07,1.07,0.00,1.07 \\ 
 s3\_405\_27270.cnfU & 405 & 27270 & 0.0074 & 202 & 1.00 & 335 & 1972.47 & 349,350,350.75,2.02,357 & 1.04,1.04,1.05,0.01,1.07 \\ 
 s3\_729\_88452.cnfU & 729 & 88452 & 0.0041 & 364 & 1.00 & 617 & 3995.53 & 665,665,665,0,665 & 1.08,1.08,1.08,0.00,1.08 \\ 

\midrule

{\bf orlib}     &      &     &  &  &    &       &                    &\\ 
 $**$scpb1.cnfU & 3000 & 300 & 0.0499 &  29 & 0.10 & 22 & 87.16 & 22,24,23.93,0.5,25 & 1.00,1.09,1.09,0.02,1.14 \\ 
 $**$scpc1.cnfU & 4000 & 400 & 0.0200 &  21 & 0.10 & 44 & 160.4 & 44,47,46.86,0.82,50 & 1.00,1.07,1.06,0.02,1.14 \\ 
 $**$scpd1.cnfU & 4000 & 400 & 0.0501 &  39 & 0.10 & 25 & 106.34 & 25,27,26.67,0.48,28 & 1.00,1.08,1.07,0.02,1.12 \\ 
 $**$scpb1.cnfW & 3000 & 300 & 0.0499 &  29 & 0.10 & 69 & 273.35 & 72,76,75.73,2.06,85 & 1.04,1.10,1.10,0.03,1.23 \\ 
 $**$scpc1.cnfW & 4000 & 400 & 0.0200 &  21 & 0.10 & 227 & 827.5 & 249,257,256.67,2.83,265 & 1.10,1.13,1.13,0.01,1.17 \\ 
 $**$scpd1.cnfW & 4000 & 400 & 0.0501 &  39 & 0.10 & 60 & 255.21 & 66,71,70.9,1.66,78 & 1.10,1.18,1.18,0.03,1.30 \\
 scp41.cnfW & 1000 & 200 & 0.0200 &  11 & 0.20 & 429 & 1295.53 & 461,463,466.94,5.1,473 & 1.07,1.08,1.09,0.01,1.10 \\ 
 scp42.cnfW & 1000 & 200 & 0.0199 &  10 & 0.20 & 512 & 1499.63 & 568,580,582.46,9.85,612 & 1.11,1.13,1.14,0.02,1.20 \\ 
 scp43.cnfW & 1000 & 200 & 0.0199 &  11 & 0.20 & 516 & 1558.26 & 589,591,592.85,3.62,598 & 1.14,1.15,1.15,0.01,1.16 \\ 
 scp44.cnfW & 1000 & 200 & 0.0200 &  10 & 0.20 & 494 & 1446.91 & 540,547,547.8,4.18,555 & 1.09,1.11,1.11,0.01,1.12 \\ 
 scp45.cnfW & 1000 & 200 & 0.0197 &  11 & 0.20 & 512 & 1546.18 & 571,577,574,3,577 & 1.12,1.13,1.12,0.01,1.13 \\ 
 scp46.cnfW & 1000 & 200 & 0.0204 &  10 & 0.20 & 560 & 1640.22 & 603,612,611.6,5.12,620 & 1.08,1.09,1.09,0.01,1.11 \\ 
 scp47.cnfW & 1000 & 200 & 0.0196 &  12 & 0.20 & 430 & 1334.38 & 474,474,474.96,1,476 & 1.10,1.10,1.10,0.00,1.11 \\ 
 scp48.cnfW & 1000 & 200 & 0.0201 &  10 & 0.20 & 492 & 1441.05 & 521,538,538.29,8.93,557 & 1.06,1.09,1.09,0.02,1.13 \\ 
 scp49.cnfW & 1000 & 200 & 0.0198 &  11 & 0.20 & 641 & 1935.74 & 741,747,745.5,3.33,750 & 1.16,1.17,1.16,0.01,1.17 \\ 
 scp51.cnfW & 2000 & 200 & 0.0200 &  10 & 0.10 & 253 & 741.03 & 282,291,290.33,2.28,295 & 1.11,1.15,1.15,0.01,1.17 \\ 
 scp61.cnfW & 1000 & 200 & 0.0492 &  20 & 0.20 & 138 & 496.49 & 152,157,157.1,2.01,163 & 1.10,1.14,1.14,0.01,1.18 \\ 
 scpa1.cnfW & 3000 & 300 & 0.0201 &  17 & 0.10 & 253 & 870.21 & 273,286,286.03,4.68,297 & 1.08,1.13,1.13,0.02,1.17 \\

 \midrule
 {\bf random} &  &  &  &  & & & \\ 
 m100\_50\_10\_10.cnfU & 50 & 100 & 0.2000 &  31 & 1.00 & 8 & 32.22 & 8,8,8.35,0.48,9 & 1.00,1.00,1.04,0.06,1.12 \\
m100\_100\_10\_10.cnfU & 100 & 100 & 0.1000 &  17 & 1.00 & 12 & 41.27 & 13,14,13.6,0.51,15 & 1.08,1.17,1.13,0.04,1.25 \\ 
 m100\_100\_10\_15.cnfU & 100 & 100 & 0.1239 &  22 & 1.00 & 10 & 36.91 & 10,11,11.25,0.51,13 & 1.00,1.10,1.12,0.05,1.30 \\ 
 m100\_100\_10\_30.cnfU & 100 & 100 & 0.1968 &  32 & 1.00 & 9 & 36.53 & 9,10,9.95,0.83,12 & 1.00,1.11,1.11,0.09,1.33 \\ 
 m100\_100\_30\_30.cnfU & 100 & 100 & 0.3000 &  43 & 1.00 & 6 & 26.1 & 6,6,6,0,6 & 1.00,1.00,1.00,0.00,1.00 \\  
 m200\_100\_10\_30.cnfU & 100 & 200 & 0.1974 &  51 & 1.00 & 11 & 49.71 & 11,12,11.98,0.28,13 & 1.00,1.09,1.09,0.03,1.18 \\ 
 m200\_100\_30\_50.cnfU & 100 & 200 & 0.3970 &  99 & 1.00 & 6 & 31.06 & 6,6,6,0,6 & 1.00,1.00,1.00,0.00,1.00 \\                 
  \midrule
 {\bf tiny} &  &  &  &  &  &\\ 
 chvatal\_6\_5.cnfW & 6 & 5 & 0.3333 & 5 & 0.83 & 1.1 & 2.51 & 2.28,2.28,2.28,0,2.28 & 2.07,2.07,2.07,0.00,2.07 \\  
 school\_9\_11\_\_0.cnfU & 9 & 11 & 0.2424 &  3 & 1.00 & 4 & 7.33 & 4,5,4.85,0.68,6 & 1.00,1.25,1.21,0.17,1.50 \\ 
 school\_9\_16.cnfU & 9 & 16 & 0.2500 &  6 & 1.00 & 5 & 12.25 & 6,6,6,0,6 & 1.20,1.20,1.20,0.00,1.20 \\ 
 school\_9\_16.cnfW & 9 & 16 & 0.2500 &  6 & 1.00 & 10 & 24.5 & 10,10.5,10.57,0.54,11.5 & 1.00,1.05,1.06,0.05,1.15 \\ 
 school\_19\_20.cnfW & 19 & 20 & 0.1421 &  6 & 1.00 & 11.5 & 28.18 & 11.5,13.5,13.39,0.91,15.5 & 1.00,1.17,1.16,0.08,1.35 \\ 
    \midrule
    instance & nCols & mRows & mDens & mCD & mP & BKV & UB & value\_Chvatal\_stats  & BKV\_ratio\_stats\\                   
    \bottomrule
    
  \end{tabular}

  }
  

\end{table*}


Table~\ref{tb_bgmc_data} introduces all instances
we use in experiments that evaluate the performance of
the maximum matching solvers and 
the minimum set cover solvers.
Columns that characterize each instance,
both for the maximum matching problem {\it as well as} for
the minimum set cover problem include:
the number of instance columns ({\it nCols}), 
the number of instance rows ({\it mRows}), 
the matrix density column  
({\it mDens $:=$ numEdges$/$(nCols*mRows)}), and
the maximum matrix degree column  ({\it mCD}).
Only the column {\it mP} relates to  
the maximum matching problem: it
denotes the percentage of columns that form the maximum matching
({\it mP $:=$ max\_matching$/$nCols}).
The remainder of columns, starting with the 
best-known-value of the minimum set cover ({\it BKV}), will be explained
in next section. All datasets and programs to support
replications of results in this paper are available
at~\cite{OPUS-github-rBedPlus-bgmc}.

\begin{figure*}[t!]
\vspace*{-3ex}
\centering
\includegraphics[width=1.00\textwidth]{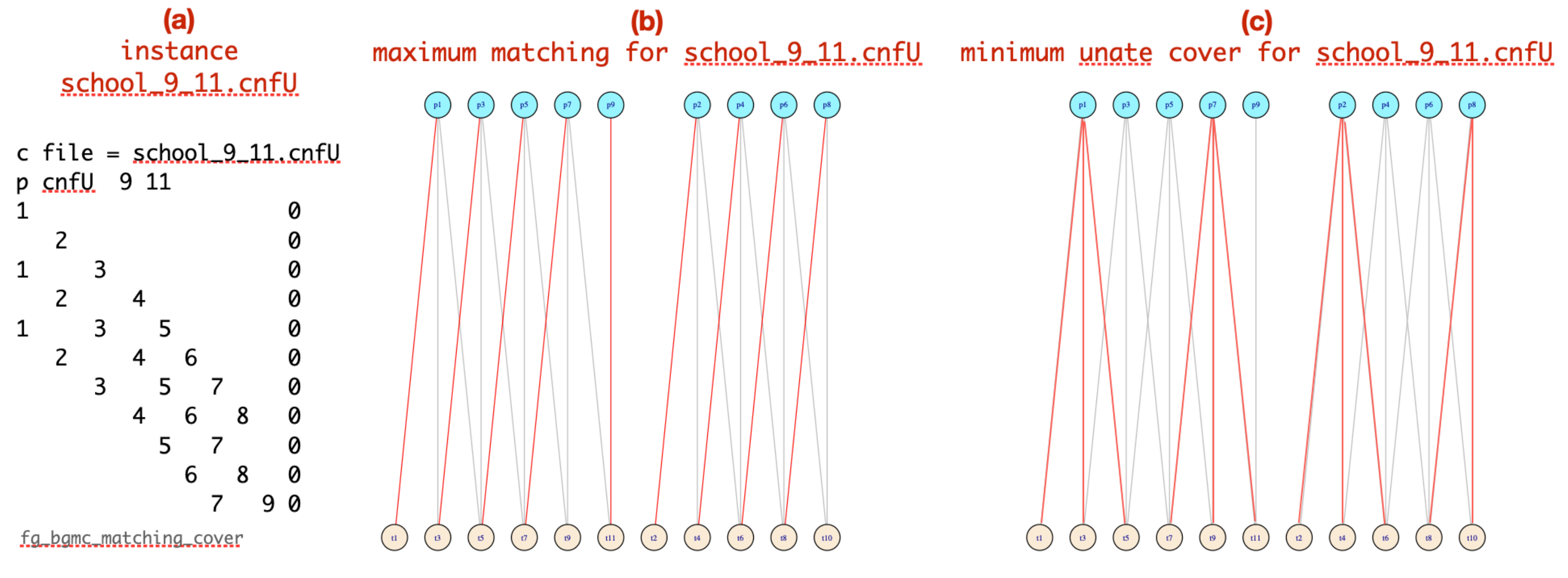}

\caption{
Three views of the
instance  {\tt school\_9\_11\_\_0.cnfU} 
introduced in Table~\ref{tb_bgmc_data}: 
{\sf(a)}~
an 11-row, 9-column  sparse matrix 
in a {\it cnf format}~\cite{OPUS-cnf-2021-wiki},
{\sf(b)}~
the maximum bipartite matching problem:
with 9 red edges representing the optimum solution,
{\sf(c)}~
the minimum set covering problem:
4 vertices at the top cover all 11 vertices at the bottom
as illustrated with 11 edges.
}
\label{fg_bgmc_matching_cover}
\end{figure*}

\begin{figure*}[h!]
\vspace*{2ex}
\centering
\includegraphics[width=1.00\textwidth]{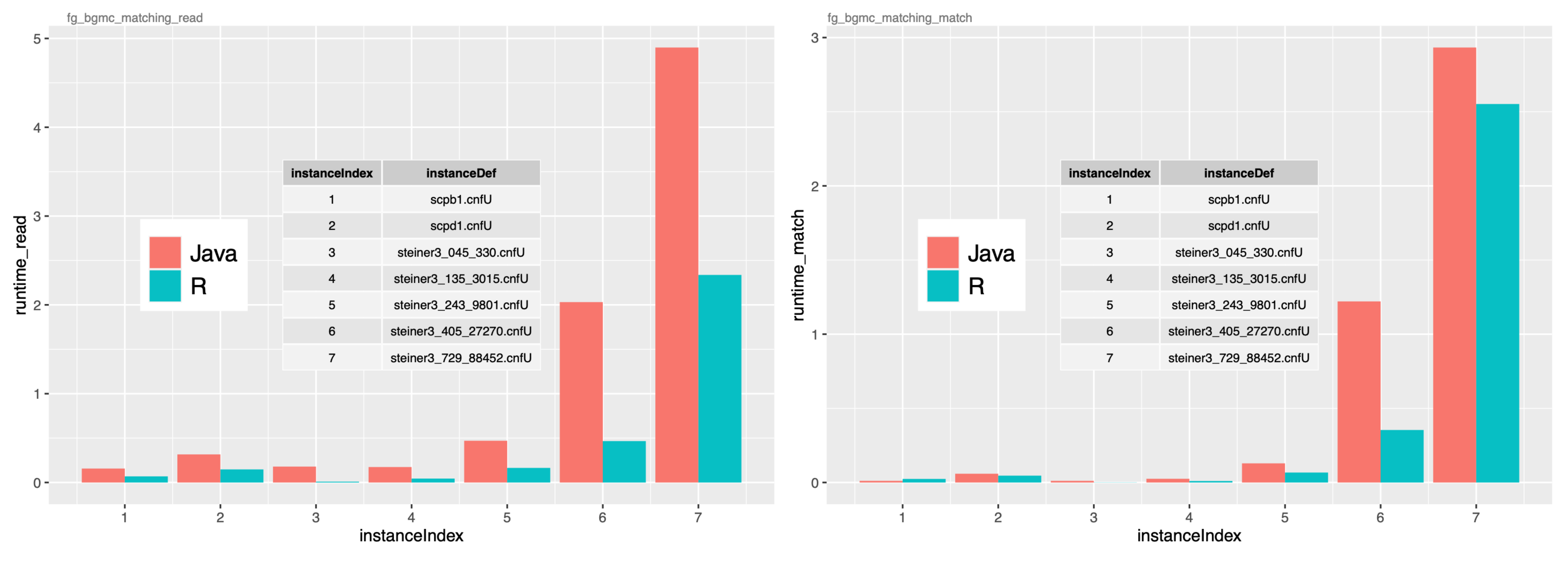}

\caption{
Maximum bipartite matching experiments with three datasets 
steiner3~\cite{OPUS-setc-2021-Resende-steiner3_data}, 
orlib~\cite{OPUS-setc-2014-orlib-Beasley}, 
and random~\cite{OPUS2-1993-benchm-Logic_synthesis}
and two solvers:
the solver in Java~\cite{OPUS-matching-2021-geeksforgeeks-java} and
the solver in \R{}~\cite{OPUS-matching-2021-igraph-R}.
The plot on left
shows the {\tt runtime\_read} for both solvers. 
The plot on right
shows the {\tt runtime\_match} for both solvers, 
i.e. the runtime to find the maximum matching. 
Only instances with runtimes $\ge$ 0.15 seconds are shown.}
\label{fg_bgmc_matching_experiment}
\end{figure*}

The example in Figure~\ref{fg_bgmc_matching_cover} illustrates 
three views of the
instance  {\tt school\_9\_11\_\_0.cnfU} 
introduced in Table~\ref{tb_bgmc_data}: 

\begin{description}

\item[\sf{Figure~\ref{fg_bgmc_matching_cover}a}]~\\\
an 11-row, 9-column  
matrix in a {\it cnf format}~\cite{OPUS-cnf-2021-wiki}.

\item[\sf{Figure~\ref{fg_bgmc_matching_cover}b}]~\\\
a bigraph as a two-layered graph 
that illustrates the {\em maximum matching problem}:
11 applicants applying for  1, 2, or 3 of the
9 jobs (teaching positions)  advertised by a school.
Each job opening can only accept one applicant and a 
job applicant can be appointed for only one job. 
In this example, 9 applicants have been matched to 9 jobs:
each match is represented by a red-colored edge.

\item[\sf{Figure~\ref{fg_bgmc_matching_cover}c}]~\\\
a bigraph as a two-layered graph 
illustrates the  {\it a unate covering problem}:
11 subjects (math, physics, etc)  can be taught by 9 instructors.
Seven instructors can teach up to 3 subjects, one instructor can
teach 2 subjects, one instructor can teach 1 subject only.
The objective of the school principal is to hire the minimum
number of teachers while still able to offer classes 
for the 11 subjects. In contrast to the maximum matching problem,
the minimum cost solution for this
covering problem is not as obvious as it is for the matching problem,
even for this small example.
There are only two minimum cost solutions: a total of  4
instructors can teach all subjects. The red-colored edges 
identify 3 instructors who will teach three
subjects and 1 instructor will teach two subjects.

\end{description}

The extension of the unate set cover to the binate set cover
problem requires addition of {\it binate clauses} as
additional rows in the sparse matrix configuration.
For example, if applicants '2' and '5' are a married couple,
and the school principal would like to hire them both,
the matrix in Figure~\ref{fg_bgmc_matching_cover}a 
will be extended with these two rows:
\\[2.5ex]
\hspace*{14ex}{\tt -2 ~~~5}\\
\hspace*{14ex}{\tt ~2 ~~-5}

On the other hand,  if applicants '4' and '7' are a divorced couple,
the school principal may prefer to find a minimum cover solution that
precludes the hiring of these two individuals together: either 
'4' or '7' may be hired but not both. In this case, 
the matrix in Figure~\ref{fg_bgmc_matching_cover}a 
will be extended with this row:
\\[1.5ex]
\hspace*{14ex}{\tt -4 ~~-7}

\subsection{{\sf Runtime experiments: Java vs R}}
\noindent
Our asymptotic experiments have been performed
with two solvers:
one implemented in Java~\cite{OPUS-matching-2021-geeksforgeeks-java}, 
the other implemented in \R{}~\cite{OPUS-matching-2021-igraph-R}.
Both rely on Ford-Fulkerson method~\cite{OPUS-matching-1956-Math-Ford_Fulkerson-max_flows}. The instances tested by both solvers have been introduced in
Table~\ref{tb_bgmc_data}. 
The results are summarized
in Figure~\ref{fg_bgmc_matching_experiment}, but only for instances with runtimes $\ge$ 0.15 seconds.
Most importantly, we separate the total runtime into two components:
(1) runtime to read and set-up all data structures ({\tt runtime\_read}),
(2) runtime to find the maximum matching ({\tt runtime\_match}).

\begin{description}

\item[\sf{runtime\_read}:]~\\\
 Java is significantly outperformed by \R{}.
 For the largest instance steiner3\_729\_88452.cnfU (729 columns, 88452 rows),
 {\tt runtime\_read\_java} $\approx$ 4.9 seconds, 
 {\tt runtime\_read\_R}    $\approx$ 2.3 seconds.
As instance size increases,
 \R{} gains  advantage when using its {\tt data.table} structure.
 In contrast, Java may need to scan each line and convert the data into a matrix.

\item[\sf{runtime\_match}:]~\\\
 All except two instances from
 the subset of OR-library instances, scpb1 and scpd1,
 are below the runtime threshold of {\it less than 0.15 seconds}.
 While Java is consistently outperformed by \R{},
 we would need larger instances to assess whether this trend holds.
 So far, the increase in {\tt runtime\_match} is monotonically increasing
 with the decreasing matrix density, both for Java and R.
 For the largest instance, steiner3\_729\_88452.cnfU,  
 {\tt runtime\_match\_java} $\approx$ 2.9 seconds while 
 {\tt runtime\_match\_R}    $\approx$ 2.5 seconds.

%

%

\OMIT{ 
 Another important observation is how the number of rows affects the runtime performance in maximum matching. By looking at the table~\ref{tb_bgmc_data}, the number of columns for scpb1.cnfU is 3000, but the number of rows is 300. The runtime performance for this instance is statistically equal to steiner3\_045\_330.cnfU, which also has around 300 rows but only has 45 columns. If we compare the runtime performance of scpb1.cnfU with steiner3\_729\_88452.cnfU, which has 88452 rows, the difference is increasing significantly.
}

\end{description}

\section{Greedy Heuristic Distributions for Set Cover} \label{sec_cover}
\noindent

\noindent
Minimum set covering problems arise in a number of domains. In logistics, the context includes market analysis, crew scheduling, emergency services, etc. Electronic design automation 
deals with logic minimization, technology mapping, and FSM optimization. 
In bioinformatics, combining Chromatin ImmunoPrecipitation (ChIP) with DNA sequencing to identify the binding sites of DNA-associated proteins
leads to formulation of the {\it motif selection problem}, mapped to a variant of the set cover problem.

An Integer Linear Programming (ILP) problem formulation
guarantees an optimum solution -- provided the solver does not time out for large problem instances. The companion 
article~\cite{OPUS2-2022-mclass-arxiv-Brglez} addresses these problems
by way of alternative stochastic approaches that go beyond the simple stochastic solver introduced in this section. Our solver is a extension of the greedy set cover algorithm by 
Chvatal~\cite{
  OPUS-setc-1979-OR-Chvatal-greedy,
  OPUS-setc-2016-Springer-Young-Greedy}.
This version, implemented in R, can significantly outperform a state-of-the-art stochastic solver in C++ for sufficiently large problem instances. 
The next three sections summarize problem isomorphs,  the  algorithm implementation, and experimental results.

\subsection{{\sf On Impact of Problem Isomorphs}}
\noindent
The idea of problem isomorphs 
to design and evaluate {\em learning experiments}
~\cite{OPUS-isomorph-1976-CognitivePhychology-Simon_Hayes}
is an on-going area of research
\cite{OPUS-isomorph-2001-Erlbaum-Gunzelmann-ACT-R_model};
it goes back to 1969
as per quote on page 382~\cite{OPUS-isomorph-1996-MITPress-Simon-Models}:
\begin{quote}
  {\it
  “I think I invented the idea of problem isomorphs about 1969 
   or a little earlier … as a follow-up on the AI researcher 
   Saul Amarel’s comment that the 
   \underline{representation of the problem} 
   could sometimes greatly facilitate its solution …”
   }
\end{quote}

\begin{figure}[t!]
\vspace*{-2ex}
\centering
\includegraphics[width=0.45\textwidth]{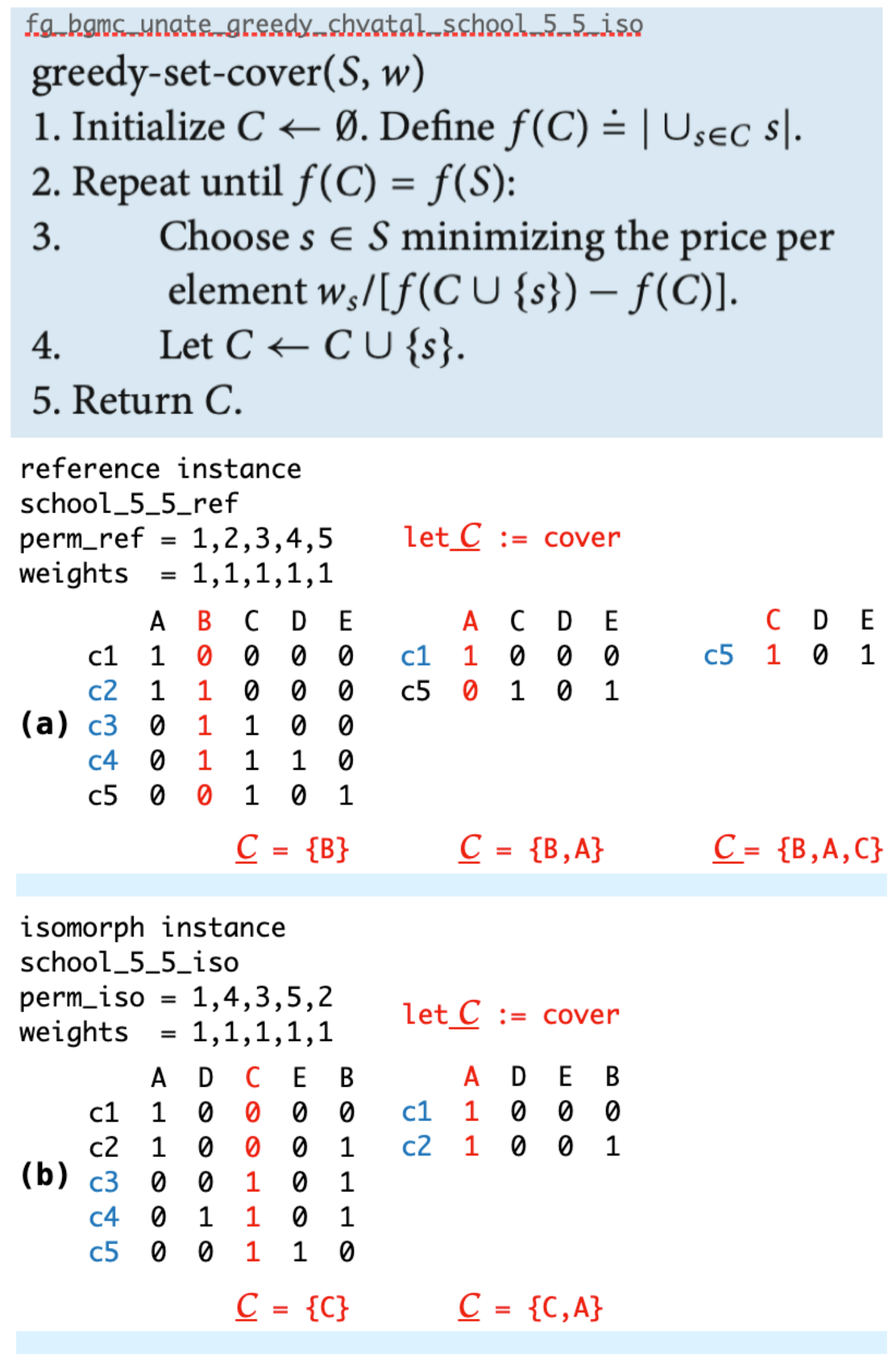}
\caption{
The pseudo code of the greedy algorithm above is 
from~\cite{OPUS-setc-2016-Springer-Young-Greedy};
it represents the version of the Chvatal's algorithm 
in~\cite{OPUS-setc-1979-OR-Chvatal-greedy}.
The R-function that implements this algorithm is named
{\tt unate\_greedy\_chvatal\_basic} 
in Figure~\ref{fg_bgmc_unate_greedy_chvatal_stoc}a.
The algorithm is invoked on two instance bigraphs, represented
as two incidence matrices:
(a) the reference instance {\tt school\_5\_5\_ref}, and 
(b) the instance isomorph  {\tt school\_5\_5\_iso}.
As implied by the matrix structure, the weight of each column is 1.
For (a), the algorithm selects the 
{\it first column with minimum rate between column weight and column degree},
i.e. column B. In the next iteration, there remain only two rows to consider: s1 and s5. The first column with maximum degree is now column A. The algorithm terminates by selecting column C, returning a solution as 3 columns \{B, A, C\} and the total weight of 3.
For (b), the algorithm selects C as the first column with minimum rate between column weight and column degree. The next iteration is also the last. With the selection of
column A,  we get a solution as 2 columns \{C, A\} and the total weight of 2.
\vspace*{-4ex}
}
\label{fg_bgmc_unate_greedy_chvatal_iso_school_5_5}
\end{figure}

The isomorphs revisited in this article have a different context
and formulation. Their merits, to change the
{\it representation} of the problem without changing the problem itself,
have already been demonstrated
by solving instances of combinatorial problems. 
In the worst case,  solvers may timeout for some, but not all, isomorphs before finding the optimum solution 
\cite{
OPUS2-2001-BDD-STTT-Harlow,
OPUS2-2001-crossnum-JEA-Stallmann,
OPUS2-2005-sat-AMAI-Brglez,
OPUS2-2005-cover-DAC-Li,
OPUS2-2007-isomorph-ExpCS-Brglez}.
\par
The pseudo code of the greedy algorithm in 
Figure~\ref{fg_bgmc_unate_greedy_chvatal_iso_school_5_5},
is the core for our stochastic algorithm.
Two instance isomorphs illustrate the  
importance of  {\it representation}
when invoking the same greedy algorithm on each of the two
instances: we observe two solutions that differ by 50\%.
A narative that follows provides a simple interpretation
on how these two instances could have been created:
\begin{quote}
{\small
Scale down the bigraph instance in Figure~\ref{fg_bgmc_matching_cover}
and transform it to an incidence matrix with 5 columns and 5 rows.
The 5 columns represent 5 applicants
\{A, B, C, D, E\} who applied to teach one or more
of the 5 classes \{c1, c2, c3, c4, c5\}.
Two administrators are performing interviews with all applicants.
Administrator (a) interviews applicants in alphabetical order and marks
course qualifications for each applicant. 
Applicant B, under the second column, is qualified to  teach classes \{c2, c3, c4\}.
Administrator (b) interviews applicants in a permuted order, \{A, D, C, E, B\}, and marks 
qualification for each applicant in the incidence matrix (b): 
marks about applicant B are now entered into the column 5.
}
\end{quote}

\par\noindent
The performance of  many greedy algorithms is measured as
a {\it ratio}:
\par\vspace*{-2ex}
\begin{equation}   \label{eq_ratio}
  {\it  ratio} := {\it  value\_greedy} / {\it  BKV}
\end{equation}
where {\it value\_greedy} is returned by the greedy algorithm 
and {\it BKV} is the {\it best-known-value} (BKV),
associated with the given instance such as 
{\tt school\_5\_5\_ref}.
Ideally, BKV represents the
proven optimum solution with an ILP-like solver,
otherwise we use the best known published value. 
As expected for the  example above,
BKV = 2 for both {\tt school\_5\_5\_ref} and {\tt school\_5\_5\_iso}.

The complete R-code of the stochastic greedy algorithm 
that relies on invoking any number of instance isomorphs is
depicted in 
Figure~\ref{fg_bgmc_unate_greedy_chvatal_stoc}a.
Consistent with definition in Eq.~\ref{eq_ratio}, 
Table~\ref{tb_bgmc_unate_greedy_chvatal_school_5_5_iso}
reports a statistical summary of experiments that involve
100 isomorphs of {\tt school\_5\_5\_ref} and 
100 isomorphs of {\tt school\_5\_5\_iso}.
\begin{table}[h!]
\vspace*{-1.1ex} 

\caption{
A statistical summary of results, based on  experiments that involve
100 isomorphs, initialized with 100 seeds, created from each of the two instances: 
{\tt school\_5\_5\_ref} and {\tt school\_5\_5\_iso}.
For the complete R-code of the stochastic greedy algorithm 
that relies on invoking any number of instance isomorphs, the algorithm
{\tt unate\_greedy\_chvatal\_iso} in
Figure~\ref{fg_bgmc_unate_greedy_chvatal_stoc}a.
However, identical results can also be generated,
using the same seeds,
with the alternative algorithm 
{\tt unate\_greedy\_chvatal\_stoc} in
Figure~\ref{fg_bgmc_unate_greedy_chvatal_stoc}b.
}
\hspace*{0.5em}
\vspace*{-1ex}
\begin{minipage}{0.40\textwidth}
\begin{Verbatim}[frame=lines, fontsize=\footnotesize,numbers=left,
numbersep=3pt,firstline=1,xleftmargin=9mm]
instanceDef = school_5_5_ref.cnfU 
unate_greedy_chvatal_iso_experiment_distr(
                              instanceDef)
                                                            
num_seeds =    100 ;    1,000  ;    10,000
  ratio  ratio_cnt ; ratio_cnt ; ratio_cnt
    1.0         45        541         5013
    1.5         55        459         4987
%
instanceDef = school_5_5_iso.cnfU 
unate_greedy_chvatal_iso_experiment_distr(
                              instanceDef)
                                                            
num_seeds =    100 ;    1,000  ;    10,000
  ratio  ratio_cnt ; ratio_cnt ; ratio_cnt
    1.0         55        459         4987
    1.5         45        541         5013
\end{Verbatim}
\end{minipage}

\label{tb_bgmc_unate_greedy_chvatal_school_5_5_iso}
\end{table}
\par
The main conclusion of the experiment in 
Table~\ref{tb_bgmc_unate_greedy_chvatal_school_5_5_iso} is this:
for both isomorph classes, with increasing number of seeds
\{100, 1,000, 10,000\} for
{\tt school\_5\_5\_ref} and {\tt school\_5\_5\_iso},
the probabilities of the best-case {\tt ratio} = 1.0 
and the worst-case {\tt ratio} = 1.5,  
converge to 0.50\%. This results is specific for the
instances which are isomorphs themselves. 
For a divers distribution of
{\tt ratio} spreads, see the results in
Figure~\ref{fg_bgmc_unate_greedy_chvatal_stoc_distr}
and
Table~\ref{tb_bgmc_data}.
 
\begin{figure}[t!]
\vspace*{-2ex}
\centering
\includegraphics[width=0.45\textwidth]{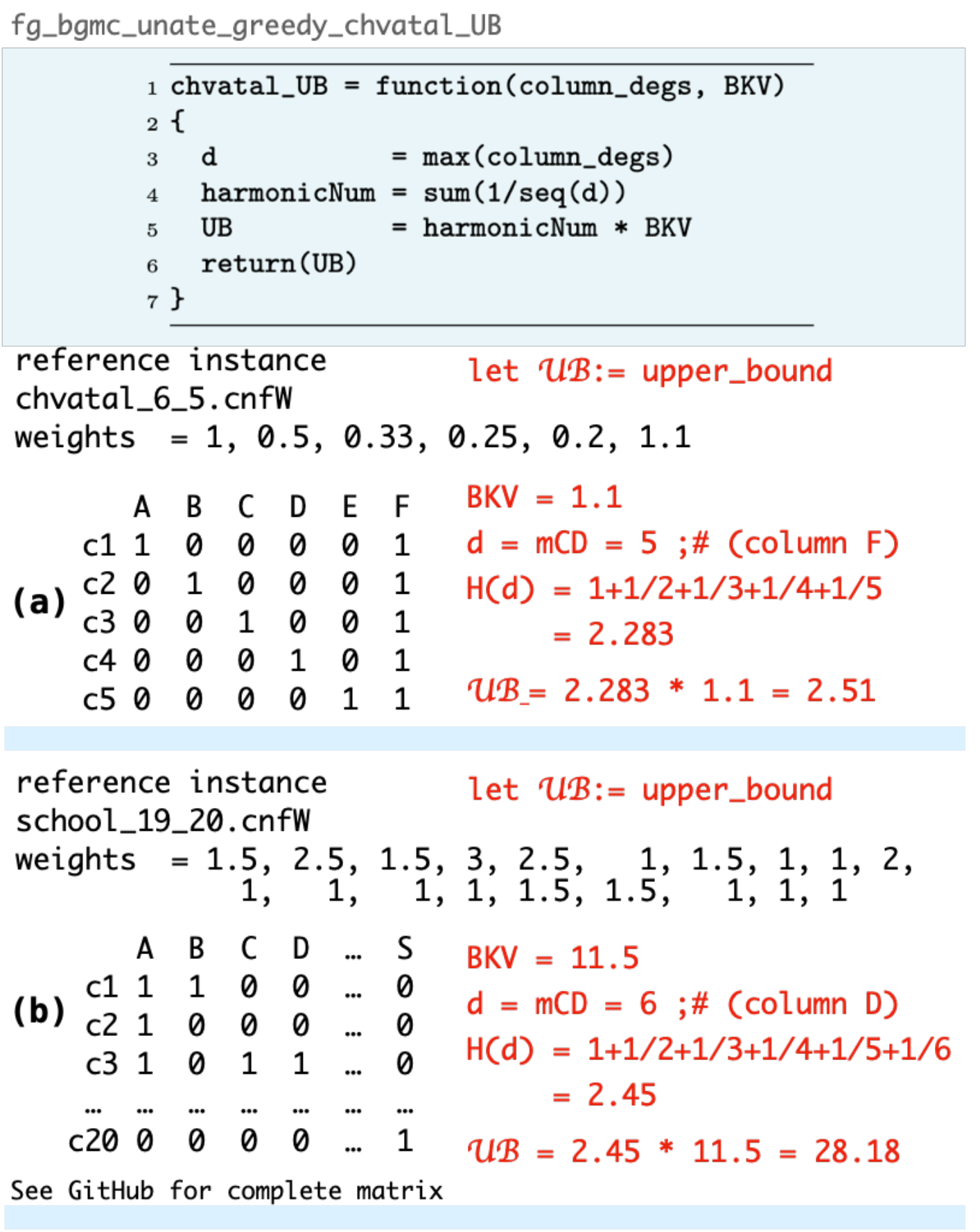}
\caption{
Two examples of computing the minimum set cover upper bound UB.
The first instance is a test case introduced 
in~\cite{OPUS-setc-1979-OR-Chvatal-greedy}.
\vspace*{-2.5ex}
}
\label{fg_bgmc_unate_greedy_chvatal_UB}
\end{figure}

Related to the worst-case ratio reported by the greedy algorithm
is the upper bound {\it UB} on the maximum value that could be
returned by the greedy algorithm. 
The search for  `tight'  upper bounds 
is still ongoing,
e.g.~\cite{
OPUS-setc-2012-LBCS-Saket-UB,
OPUS-setc-2016-OR-Felici-UB,
OPUS-setc-2019-STOC-Abboud-UB}.
%
%
%
%
However, not one of these publications offers
{\em empirical evidence} of how tight these bounds really are
for {\em any} specific instances
relatively to Chvatal's bound in~\cite{OPUS-setc-1979-OR-Chvatal-greedy}.
The upper bound UB we use in 
this article has been formulated in~\cite{OPUS-setc-1979-OR-Chvatal-greedy}.
For an illustration of how we apply this bound to instances in this article,
see Figure~\ref{fg_bgmc_unate_greedy_chvatal_UB}.

The next section 
introduces a simplification of the greedy algorithm, 
replacing the isomorph-based solver  
{\tt unate\_greedy\_chvatal\_iso}
with an alternative solver,
{\tt unate\_greedy\_chvatal\_stoc}.

\begin{figure*}[t!]
\vspace*{-9ex} 
\hspace*{-2.1em}
\begin{minipage}{0.51\textwidth}
\centering

{\large\bf  (a) }
\vspace*{-1ex} 
\begin{Verbatim}[frame=lines, fontsize=\footnotesize,numbers=left,
numbersep=3pt,firstline=1,xleftmargin=9mm]
unate_greedy_chvatal_basic = function() 
{
  # required inputs
  n           = glob[["nCols"]]
  m           = glob[["mRows"]]
  M           = glob[["M_ref"]]
  colWeights  = glob[["colWeights_ref"]]  
  # local initializations
  nOps        = 0
  coord       = rep(0, n)
  
  while(TRUE) {
    percentages = colWeights / colSums(M) 
    if (all(percentages == Inf)) { break }
    jdx         = which.min(percentages)
    rem_vec     = which(M[,jdx] %in% 1)
    M[rem_vec,] = 0
    coord[jdx]  = 1
    nOps        = nOps + 1
  }
  coordGreedy = paste(coord, collapse = "")
  valueGreedy = as.numeric(t(coord) %*% colWeights))
  
  return(list(
    coordGreedy = coordGreedy, 
    valueGreedy = valueGreedy,
    nOps        = nOps))
} 
\end{Verbatim}
\vspace*{-6.5ex}
\begin{Verbatim}[frame=lines, fontsize=\footnotesize,numbers=left,
numbersep=3pt,firstline=1,xleftmargin=9mm]
unate_greedy_chvatal_iso = function(replicaId=0) 
{  
  # required inputs 
  n               = glob[["nCols"]]
  m               = glob[["mRows"]]
  M_ref           = glob[["M_ref"]]
  colWeights_ref  = glob[["colWeights_ref"]]
  greedyId        = glob[["greedyId"]]   
  if (replicaId == 0) {
    coordPermV = 1:n # reference permutation (natural order)
    coordPerm  = paste(coordPermV, collapse=",")
    colWeights = glob[["colWeights_ref"]] 
    M          = glob[["M_ref"]]
 
  } else {
    # create an isomorph instance, controlled by replicaId
    set.seed(replicaId)
    coordPermV = sample(1:n)
    coordPerm  = paste(coordPermV, collapse=",")
    colWeights = c()
    M          = matrix(rep(NA, m*n), ncol=n)
    for (idx in 1:n) {
      i = idx
      j = coordPermV[idx] 
      colWeights[idx] = glob[["colWeights_ref"]][j]
      M[ ,idx]        = glob[["M_ref"]][,j]
    }
  }
  # invoke unate_greedy_chvatal_basic() with new variables 
  glob[["M_ref"]]          = M
  glob[["colWeights_ref"]] = colWeights
  glob[["replicaId"]]      = replicaId 
  answ = unate_greedy_chvatal_basic()
  
  coordGreedy = answ$coordGreedy
  valueGreedy = answ$valueGreedy ; nOps = answ$nOps
  return(list(coordGreedy=coordGreedy, 
              valueGreedy=valueGreedy, nOps=nOps))
 }
  \end{Verbatim}
\end{minipage}
\begin{minipage}{0.51\textwidth}
\centering

{\large\bf  (b) }
\vspace*{-1ex} 
\begin{Verbatim}[frame=lines, fontsize=\footnotesize,numbers=left,
numbersep=3pt,firstline=1,xleftmargin=9mm]
unate_greedy_chvatal_stoc = function() 
{
  # required inputs
  M          = glob[["M_ref"]]
  n          = glob[["nCols"]]
  m          = glob[["mRows"]]
  colWeights = glob[["colWeights_ref"]]
  replicaId  = glob[["replicaId"]]
  # local initializations
  nOps       = 0
  coord      = rep(0, n)
  
  while(TRUE) {
    percentages = colWeights / colSums(M)
    if (all(percentages==Inf)) { break }
    if (replicaId == 0) {
      jdx     = which.min(percentages)
    } else {
      jdx_vec = which(percentages == min(percentages))
      jdx_cnt = sample(1:length(jdx_vec))[1]
      jdx     = jdx_vec[jdx_cnt]         
    }
    rem_vec     = which(M[,jdx] %in% 1)
    M[rem_vec,] = 0
    coord[jdx]  = 1
    nOps        = nOps + 1
  }
  
  coordGreedy = paste(coord, collapse = "")
  valueGreedy = as.numeric(t(coord) %*% colWeights))
  
  return(list(
    coordGreedy = coordGreedy, 
    valueGreedy = valueGreedy,
    nOps        = nOps))
}
\end{Verbatim}
\vspace*{-6.5ex}
\begin{Verbatim}[frame=lines, fontsize=\footnotesize,numbers=left,
numbersep=3pt,firstline=1,xleftmargin=9mm]
unate_greedy_chvatal_stoc_experiments = 
  function(instanceDef, isSeedConsecutive=T, replicateSize=10) {
  
  # read instance file and convert to matrix with detailed info
  # data store in global list, glob
  
  read_bgu(instanceDef)
  dt = data.table()
  for (replicaId in 0:replicateSize) {
    
    glob[["replicaId"]] = replicaId
    if (isSeedConsecutive) {
      seedInit = replicaId 
    } else {
      seedInit = trunc(1e6*runif(1)) 
    }
    set.seed(seedInit)
    
    answ = unate_greedy_chvatal_stoc() 
    coordGreedy = answ$coordGreedy
    valueGreedy = answ$valueGreedy ; nOps = answ$nOps
    dt = rbind(dt, list(
      replicaId   = replicaId,
      nOps        = nOps,
      coordGreedy = coordGreedy,
      valueGreedy = valueGreedy
    ))
  }
  return(dt)
  
}
\end{Verbatim}

\end{minipage}
\vspace*{2ex}
\caption{
Two equivalent {\it stochastic versions} in R of the Chvatal's algorithm:   
(a) inducing distributions of set covers with bigraph isomorphs and
(b) inducing distributions of set covers by randomizing best selections. To achieve the randomization, when ${\rm replicaId} > 0$, we use the random selection returned by the R-function "which".
}
\label{fg_bgmc_unate_greedy_chvatal_stoc}
\end{figure*}

\subsection{{\sf On Set Covers with a Stochastic Greedy Algorithm}}
\noindent
The simplified stochastic greedy algorithm, 
replacing the isomorph-based implementation,
is represented by the function
{\tt unate\_greedy\_chvatal\_stoc} 
in Figure~\ref{fg_bgmc_unate_greedy_chvatal_stoc}b.
The distributions of set cover solutions are induced,
for ${\rm replicaId} > 0$, 
by relying on the randomized selections returned by the R-function "which".

How can we claim that the very different implementations 
of the two greedy stochastic algoritms in 
Figures~\ref{fg_bgmc_unate_greedy_chvatal_stoc}a and 
\ref{fg_bgmc_unate_greedy_chvatal_stoc}b are equivalent?
In Figures~\ref{fg_bgmc_unate_greedy_chvatal_stoc}a 
we induce randomization by explicitly interchanging rows
in the matrix.
In Figures~\ref{fg_bgmc_unate_greedy_chvatal_stoc}b
we induce randomization by a random selection of columns 
with the same minimum rate between column weight and column degree.

Experiments have demonstrated, for {\em both solvers}, 
the same or close to the same distributions
of ratios such as shown in 
Table~\ref{tb_bgmc_unate_greedy_chvatal_school_5_5_iso},
and
Figure~\ref{fg_bgmc_unate_greedy_chvatal_stoc_distr}.
However, the runtime of 
{\tt unate\_greedy\_chvatal\_stoc} solver that relies on changing the seed
has better  runtime and easier interpretation than
{\tt unate\_greedy\_chvatal\_iso} solver that
relies on explicit permutations of matrix columns.
For specific isomorphs of interest, 
we do invoke a column permutation of
the respective reference instance such as
the example of instance {\tt school\_5\_5\_ref}.
\subsection{{\sf Runtime Experiments}}
\begin{figure*}[t!]
\vspace*{-4ex}
\centering

{\sf (school\_9\_11.cnfU)} 
\includegraphics[width=0.90\textwidth]{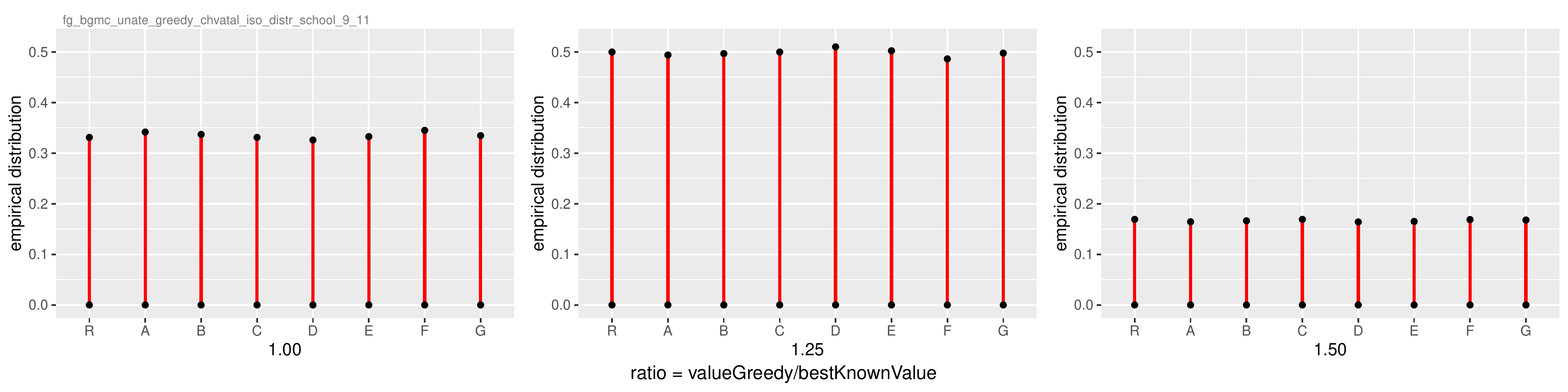}
\\[2ex]
{\sf (ab)}\\
\includegraphics[width=0.90\textwidth]{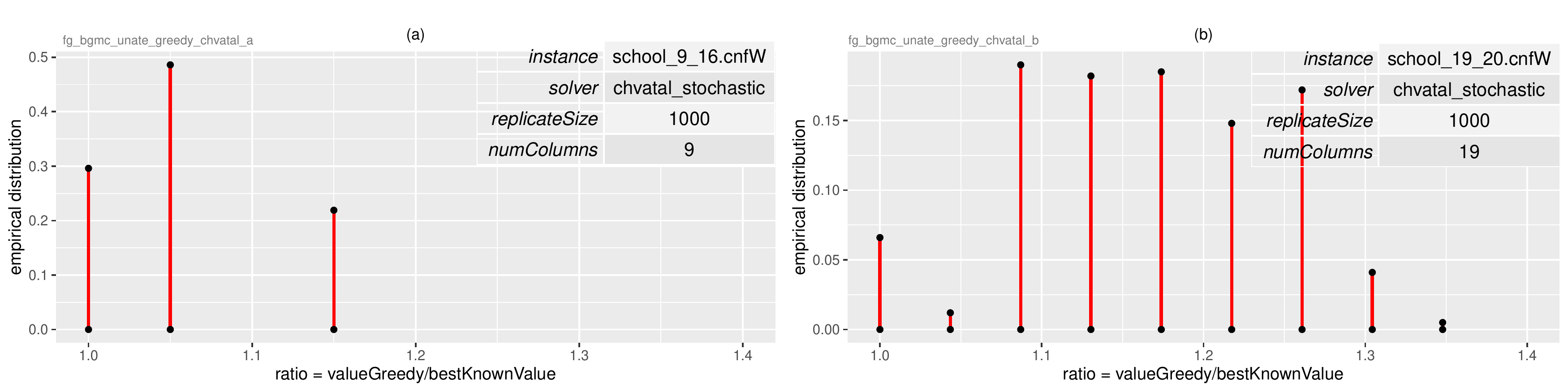}
\\[2ex]
{\sf (cd)}\\
\includegraphics[width=0.90\textwidth]{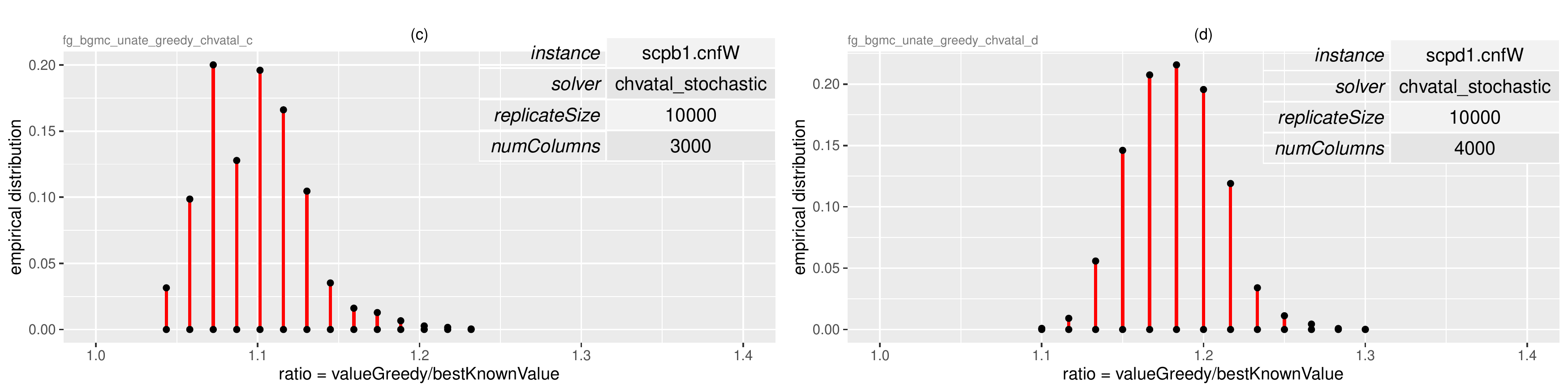}

\caption{
Reference parameters for each instance in this figure
are listed in Table~\ref{tb_bgmc_data}.
The top segment {\sf (school\_9\_11.cnfU)}
depicts  the empirical distribution of ratios \{1.0, 1.25, 1.5\}
for the 8 isomorph classes of size 10,000 each, induced by
the instance {\tt school\_9\_11.cnfU}. %
The middle segment {\sf (ab)}
depicts  distribution of ratios  
for two isomorph classes, each of size 1,000.
The bottom segment {\sf (cd)}
depicts  distribution of ratios the 
for two isomorph classes, each of size 10,000.
For additional details, see the article.
\vspace*{-3ex}
}
\label{fg_bgmc_unate_greedy_chvatal_stoc_distr}
\end{figure*}

\noindent
The runtime experiments with the minimum set cover instances are summarized in
Figure~\ref{fg_bgmc_unate_greedy_chvatal_stoc_distr},
Table~\ref{tb_bgmc_data}, 
Table~\ref{tb_bgmc_school_19_20}, and
Table~\ref{tb_bgmc_data_BRKGA}.

\begin{description}

\item[\sf{Figure~\ref{fg_bgmc_unate_greedy_chvatal_stoc_distr}}]~\\\
The top segment {\sf (school\_9\_11.cnfU)}
depicts  the empirical distribution of ratios \{1.0, 1.25, 1.5\}
for the 8 isomorph classes of the instance {\tt school\_9\_11.cnfU}.
This instance, 
already introduced in Figure~\ref{fg_bgmc_matching_cover},
induces isomorph classes \{{\sf R, A, B, C, D, E, D,F, G}\}.
The {\it reference class} {\sf R} represents 10,000 instance isomorphs of 
{\tt school\_9\_11\_R} = {\tt school\_9\_11}.
The {\it alternative class} {\sf A} represents 10,000 instance isomorphs of 
{\tt school\_9\_11\_A} = (seed-specific isomorph of {\tt school\_9\_11}).
The remainder of classes, \{B,...,G\}, are formed similarly.
The empirical distribution probabilities associated with each ratio
of the reference isomorph class {\sf R} of size 10,000 are itemized below:  
\par\vspace*{1.2ex}
\hspace*{1.6em}
\begin{minipage}{0.33\textwidth}
\begin{Verbatim}[frame=lines, fontsize=\footnotesize,numbers=left,
numbersep=3pt,firstline=1,xleftmargin=9mm] 
   ratio ratio_cnt empirical_distr
1:  1.00      3311          0.3311
2:  1.25      4998          0.4998
3:  1.50      1691          0.1691
\end{Verbatim}
\end{minipage}

\par\vspace*{1.2ex}\noindent
As shown in the plot, empirical distributions of ratios of the
isomorph in classes from  \{A,...,G\}
closely match the 
reference isomorph class {\sf R}.
However, the variance between the isomorph classes
becomes noticeable with decreasing the
class size from 10,000 to 1,000 and 100 
-- just as already observed in
Table~\ref{tb_bgmc_unate_greedy_chvatal_school_5_5_iso}.

The middle segment labeled as {\sf (ab)}
depicts  distribution of ratios  
for two isomorph classes, each of size 1,000:
weight-specific {\tt school\_9\_16.cnfW} and 
weight-specific {\tt school\_19\_20.cnfW}.
For  the {\tt school\_9\_16.cnfW} isomorph class, 
we observe 3 distinct ratios in the range  \{1.0, 1.15\}.
For the {\tt school\_19\_20.cnfW} isomorph class,
we observe  9 distinct ratios in the range  \{1.0, 1.35\}.

The bottom segment of this figure, labeled as {\sf (cd)}
depicts  the empirical distribution of ratios  
for two isomorph classes, each of size 10,000:
weight-specific {\tt scpb1.cnfW} and 
weight-specific {\tt scpd1.cnfW}.
For the {\tt scpb1.cnfW} isomorph class,
we observe  14 distinct ratios in the range \{1.04, 1.23\}.
For the {\tt scpd1.cnfW} isomorph  class,
we observe  13 distinct ratios in the range \{1.10, 1.30\}.

\item[\sf{Table~\ref{tb_bgmc_data}}]~\\\
This table
introduces all instances
that summarize results of experiments in this article.
Details about instance parameters have been discussed
in Section~\ref{sec_matching} when introducing
the maximum matching solvers.
Additional columns in this table, relevant to
the minimum set cover solvers  
in this section include
{\it BKV, UB, value\_Chvatal\_stats, and BKV\_ratio\_stats}.
The definition of {\it ratio} in Equation~\ref{eq_ratio}
relies on the {\it best-known-value} {\it BKV}. 
The computation of the Chvatal's upper bound 
{\it UB} on the minimum set cover is illustrated in
Figure~\ref{fg_bgmc_unate_greedy_chvatal_UB}.
The column {\it value\_Chvatal\_stats} reports the 
empirical set cover statistics with solver {\tt unate\_greedy\_chvatal\_stoc},
the stochastic version of the Chvatal's greedy algorithm.
The reported statistics represents comma-separated
values of minimum, median, mean, standard deviation, and maximum.
The column {\it  BKV\_ratio\_stats} reports
the same statistics, normalized with respect to {\it BKV}.

There are 37 instances in Table~\ref{tb_bgmc_data}.
The question arises of how effective 
the stochastic greedy algorithm solver
{\tt unate\_greedy\_chvatal\_stoc} 
actually is.
A partial glimpse is shown in the table below:

\par\vspace*{1.2ex}
\hspace*{0.8em}
\begin{minipage}{0.36\textwidth}
\begin{Verbatim}[frame=lines, fontsize=\footnotesize,numbers=left,
numbersep=3pt,firstline=1,xleftmargin=9mm]      
    ranges_of_ratios  counts_out_of_37     
        ratio  = 1.0                12                
1.00 <  ratio <= 1.1                18                
1.20 <= ratio <= 2.5                 7                 
\end{Verbatim}
\end{minipage}
\par\vspace*{1.1ex}\noindent
The counts about in the table above signify:
\begin{itemize}
\item
optimum solutions have been found for 12-out-37 instances, 
\item
solutions within 10\% of the optimium have been found for 18-out-37 instances,
\item
solutions above 20\% of the optimium have been found for 7-out-37 instances.
\end{itemize}

\begin{table*}[h!]
\vspace*{1.9ex} 
\caption{
Similar size instances and their variabilities: BKVs, UBs, Chvatal ratios, and harmonic numbers.
For  details, see the article.
}
\label{tb_bgmc_school_19_20}
\hspace*{2.8em}
\begin{minipage}{0.85\textwidth}
\begin{Verbatim}[frame=lines, fontsize=\footnotesize,numbers=left,
numbersep=3pt,firstline=1,xleftmargin=9mm] 
                                    exact                                chvatal  chvatal  
                                    column_solutions/                    best     worst   harmonic  
instance             weight_range   column_degrees        BKV      UB    ratio    ratio     number  
school_19_20.cnfU    [1.0,   1.0]   2,4,5,6,15,16/         6      14.7    1.0      1.0      2.45
                                    5,6,5,1, 3, 3
school_19_20.cnfW    [1.0,   3.0]   1,5,7,9,10,14,17,18/  11.5    28.18   1.0      1.348    2.45      
                                    3,5,3,2, 4, 2, 2, 2
school_19_20_1.cnfW  [0.333, 1.0]   1,5,7,9,10,14,17,18/   3.833   9.39   1.087    1.304    2.45
                                    3,5,3,2, 4, 2, 2, 2    
school_19_20_5.cnfW  [0.7,   4.9]   1,4,5,7,9,15,16/      16.1    39.45   1.087    1.087    2.45
                                    3,6,5,3,2  3, 3           
chvatal_19_18.cnfW   [0.33,  1.1]   19/                    0.5     1.75   1.0      1.0      3.496 
                                    18                            
\end{Verbatim}
\end{minipage}
%
%
%
%
\vspace*{-2.5ex}
\end{table*}  

\item[\sf{Table~\ref{tb_bgmc_school_19_20}}]~\\\
This table assembles four versions of 19-column, 20-row instance 
{\tt school\_19\_20}: one with {\it all weights at 1}, and three with
weights in the ranges shown.
The fifth instance, {\tt chvatal\_19\_18}, is
a scaled-up instance of {\tt chvatal\_6\_5} introduced 
in~\cite{OPUS-setc-1979-OR-Chvatal-greedy}.
In {\tt chvatal\_6\_5}, only the weight of the column 6
determines its BKV.  In {\tt chvatal\_19\_18},
BKV = 0.5 for the solution column 19 and weight = 0.5.
For more observations, see below:
\begin{itemize}
\item
As already demonstrated in Table~\ref{tb_bgmc_data},
the ratios UB/BKV $>$ 2, i.e. UB is not a tight upper bound. 
Weight range 
impacts UB significantly.
\item
Weight range also impacts significantly 
the minimum and the maximum range of ratio 
returned by the solver {\tt unate\_greedy\_chvatal\_stoc}.
The harmonic number is determined by
{\tt max(column\_degrees)} only.
The ratio {\tt harmonic\_number/worst\_ratio)} 
varies from 2.35/1.348 = 1.743 to 3.496.
\end{itemize}

\begin{table*}[h!]
\vspace*{2.5ex} 
\caption{
This is an extension of~Table~\ref{tb_bgmc_data}.
The {rows~3--5} refer to three large instances from the 
OR-library~\cite{OPUS-setc-2014-orlib-Beasley}.
The {rows~6--8} introduce {\it almost identical} instances
related to instances on rows  {rows~3--5} with the exception that now 
{\it all weights are set to 1.}
For 
details, see the article.
%
%
%
%
}
\hspace*{6.0em}
\begin{minipage}{0.73\textwidth}
\begin{Verbatim}[frame=lines, fontsize=\footnotesize,numbers=left,
numbersep=3pt,firstline=1,xleftmargin=9mm] 
                     chvatal         BRKGA      chvatal  chvatal     BRKGA    BRKGA
  instance   BKV  cover_best    cover_best    num_seeds  seconds  num_gens  seconds 
scpb1.cnfW    69          72            69       10,000    1,271       200    2,505 
scpc1.cnfW   227         249           227       10,000    2,601       200    5,274    
scpd1.cnfW    60          66            60       10,000    1,666       200    5,377  
scpb1.cnfU    22          22            25       10,000      838       200    3,078     
scpc1.cnfU    44          44            47       10,000    1,444       200    4,764   
scpd1.cnfU    25          25            27       10,000    1,120       200    5,713           
\end{Verbatim}
\end{minipage}
\label{tb_bgmc_data_BRKGA}
\vspace*{2ex}
\end{table*}

\item[\sf{Table~\ref{tb_bgmc_data_BRKGA}}]~\\\
This table supplements the results of 
the greedy set cover experiments summarized in Table~\ref{tb_bgmc_data}.
The six instances from 
Table~\ref{tb_bgmc_data} 
summarize the most important results obtained to date with
the solver {\tt unate\_greedy\_chvatal\_stoc},
running each instance with 10,000 unique seeds,
equivalent to processing 10,000  isomorphs 
of each of six instances.
\begin{itemize}
\item
The {rows~3--5} refer to three large instances from the 
OR-library~\cite{OPUS-setc-2014-orlib-Beasley}.
Experiments with these instances
verify the nominal performance of the 
\CPP~solver 
{\tt BRKGA}~\cite{
  OPUS-setc-2014-BRKGA-Resende-code,
  OPUS-setc-2014-BRKGA-Resende} 
reporting the minimum cover value found after running each instance 
with the {\it generation limit} of 200.
The best covers
returned by this solver match the BKVs reported 
elsewhere~\cite{OPUS-setc-2014-SWJ-Broderick-Bee_Colony}:
\{69, 227, 60\}. These covers dominate 
the best covers returned by
{\tt unate\_greedy\_chvatal\_stoc}: 
\{72, 249, 66\}.
\item
The {rows~6--8} introduce {\it almost identical} instances
related to instances on rows  {rows~3--5} with the exception that now 
{\it all weights are set to 1.}
There are no better covers 
than the ones reported by 
\R~solver {\tt unate\_greedy\_chvatal\_stoc} in this article:
\{22, 44, 25\}. The best covers 
returned by  \CPP~solver {\tt BRKGA} are significantly worse:
\{25, 47, 27\}. 
\end{itemize}

The results on lines 6--8 in Table~\ref{tb_bgmc_data_BRKGA} 
are new: they provide the currently {\it best-known-values} (BKVs) for the
three largest OR-instances listed on lines 6--8. 
This may well be the first time where a greedy algorithm outperforms
a state-of-the-art algorithm designed to search for optimum solutions -- 
not only in runtime but more importantly, in delivering 
significantly better solutions.
\end{description}


\vspace*{2ex}
\section{Summary and Future Work}  \label{sec_summary}
\noindent
The roots for this article have been  provided by the rather unexpected 
empirical result in December 2020, as a follow-up on the just completed 
CSC316 Java project in a junior-level course in data
structures and algorithms~\cite{OPUS-csc316-fall-2020}.
This result is summarized with two asymptotic plots, 
Figure~\ref{fg_bgmc_movieLib_runtime}a and Figure~\ref{fg_bgmc_movieLib_runtime}b.
Elements of surprise include not only the significant runtime improvements
with R-code versus the Java-code but also that these results were produced in a time frame of two weeks by the first author who was completely new to R.
However, as it frequently happens, the time required to {\it explain} a new result can be much longer than the time required 
to produce the result itself.

\begin{description}
\item[\sf{Note}]~\
\\
For all datasets, programs and asymptotic experiments with data structures
in this article, see~\cite{OPUS-github-rBedPlus-bgmc}.

\item[\sf{Summary}]~\
\begin{itemize}
\item
Our model of {\tt movieLib} in 
Figure~\ref{fg_bgmc_movieLib_data}b
is a simplified case of an {\it affiliation} bipartite graph.
For example, the first sentence
from~\cite{OPUS-affiliation-2021-Springer-Stankova-bg_classification} 
begins with
{\small\it `Many real-world large datasets correspond to bipartite graph 
data settings—think for example of users rating movies or people visiting locations'}.
\\
Data structures in R may well provide an
advantage over Java for the class of affiliation bipartite graphs.

\item
The maximum bipartite matching experiments in
Figure~\ref{fg_bgmc_matching_experiment}
consistently demonstrate the runtime advantages of R
data structures in comparison with Java.

\item
The introduction of two stochastic algorithms demonstrates
advantages of greedy heuristic for the {\it set cover problem}.
The merits of rapid prototyping these algorithms in R is apparent
by the simplicity and the readability of the code 
in Figure~\ref{fg_bgmc_unate_greedy_chvatal_stoc}.

\item
The results on lines 6--8 in Table~\ref{tb_bgmc_data_BRKGA} 
provide the currently {\it best-known-values} (BKVs) for the
three largest OR-instances listed on lines 6--8. 
By significantly outperforming
a state-of-the-art algorithm designed to search for optimum solutions -- 
not only in runtime but more importantly, in delivering 
significantly better solutions -- these greedy solutions
provide a strong basis and motivation for future work.

\end{itemize}

\item[\sf{Future Work}]~\
\begin{itemize}
\item
The current work focuses on completing two 
companion 
articles,~\cite{OPUS2-2022-coupon-arxiv-Brglez} 
and~\cite{OPUS2-2022-mclass-arxiv-Brglez}.
Both provide support and components for 
for the work in this article as well as for the
articles to follow.
\item
In the immediate future, methods that produced the results 
on lines 6--8 in Table~\ref{tb_bgmc_data_BRKGA} of this paper
will provide the basis for
new methods in stochastic combinatorial optimization. 

\end{itemize}

\end{description}

\section*{Acknowledgements}
\vspace*{-0.5ex}\noindent
A number of individuals and teams have
contributed to the evolution of this article.
We gratefully acknowledge them all.

\begin{itemize}
\item
Dr. Jason King gave permission to use and post course-related {\tt movieLib} and {\tt javaLib} 
after the completion his CSC316 Data Structures Course in Fall 2020.
\item
Dr. Barbara Adams advised Eason Li to enroll for a 3-hour credit course 
CSC499 in the Fall 2021 as a part of this project.
\item
The R-project team created and supports the \R-platform and environment~\cite{OPUS-R-manual}.
\item
Numerous volunteers continue to 
post valuable snippets of \R-code and advice on the Web, in particular
r-bloggers.com, stackoverflow.com, and geeksforgeeks.org.
\item
The team of the BRKGA algorithm project posted results of their research and experiments
\cite{OPUS-setc-2014-BRKGA-Resende-code,OPUS-setc-2014-BRKGA-Resende}.
\item
The team of the  ABC algorithm project for posted results of their research and experiments
\cite{OPUS-setc-2014-SWJ-Broderick-Bee_Colony}.
\end{itemize}



\end{document}